\documentclass[sigconf]{acmart}
\AtBeginDocument{%
  }

\copyrightyear{2026}
\acmYear{2026}
\setcopyright{cc}
\setcctype{by}
\acmBooktitle{Proceedings of the 2026 CHI Conference on Human Factors in Computing Systems (CHI '26), April 13--17, 2026, Barcelona, Spain}
\acmDOI{10.1145/3772318.3793266}
\acmConference[CHI '26]{Proceedings of the 2026 CHI Conference on Human Factors in Computing Systems}{April 13--17, 2026}{Barcelona, Spain}
\acmISBN{979-8-4007-2278-3/2026/04}

\usepackage{acmart-taps}

\begin{document}

\title{How Multimodal Large Language Models Support Access to Visual Information: A Diary Study With Blind and Low Vision People}

\author{Ricardo E. Gonzalez Penuela}
\email{reg258@cornell.edu}
\orcid{0000-0002-1344-3850}
\affiliation{%
  \institution{Cornell University}
  \city{New York}
  \state{New York}
  \country{USA}}

\author{Crescentia Jung}
\email{cj382@cornell.edu}
\orcid{0000-0003-4434-1723}
\affiliation{%
  \institution{Cornell University}
  \city{New York}
  \state{New York}
  \country{USA}
}

\author{Sharon Lin}
\email{syl55@cornell.edu}
\orcid{0009-0005-8282-6969}
\affiliation{%
  \institution{Cornell University}
  \city{New York}
  \state{New York}
  \country{USA}
}

\author{Ruiying Hu}
\email{rh692@cornell.edu}
\orcid{0009-0009-3273-3253}
\affiliation{%
  \institution{Cornell University}
  \city{New York}
  \state{New York}
  \country{USA}
}

\author{Shiri Azenkot}
\email{shiri.azenkot@cornell.edu}
\orcid{0000-0002-6701-4066}
\affiliation{%
  \institution{Cornell University}
  \city{New York}
  \state{New York}
  \country{USA}
}

\renewcommand{\shortauthors}{Gonzalez et al.}

\begin{abstract}
Multimodal large language models (MLLMs) are changing how Blind and Low Vision (BLV) people access visual information. Unlike traditional visual interpretation tools that only provide descriptions, MLLM-enabled applications offer conversational assistance, where users can ask questions to obtain goal-relevant details. However, evidence about their performance in the real-world and implications for BLV people's daily lives remains limited. To address this, we conducted a two-week diary study, where we captured 20 BLV participants' use of an MLLM-enabled visual interpretation application. Although participants rated the visual interpretations of the application as "trustworthy" (mean=3.76 out of 5, max=extremely trustworthy) and "somewhat satisfying" (mean=4.13 out of 5, max=very satisfying), the AI often produced incorrect answers (22.2\%) or abstained (10.8\%) from responding to users' requests. Our findings show that while MLLMs can improve visual interpretations' descriptive accuracy, supporting everyday use also depends on the “visual assistant” skill: behaviors for providing goal-directed, reliable assistance. We conclude by proposing the "visual assistant" skill and guidelines to help MLLM-enabled visual interpretation applications better support BLV people's access to visual information.
\end{abstract}


\begin{teaserfigure}
\centering
 \includegraphics[width=0.85\linewidth]{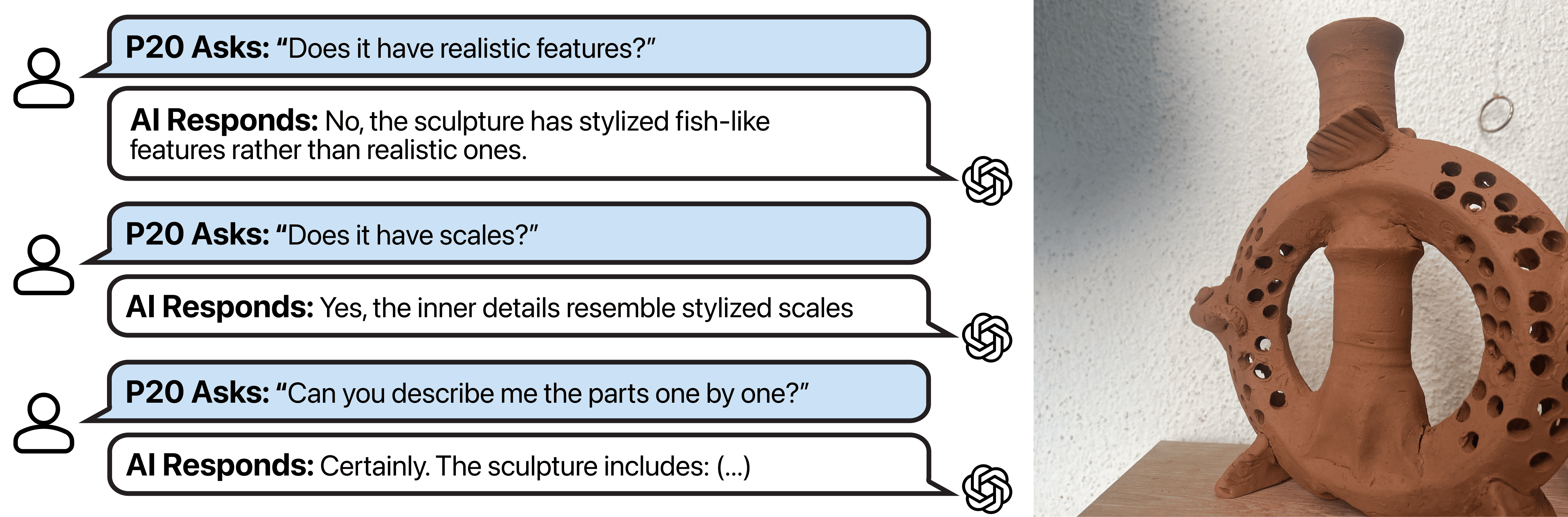}
 \Description{A Handmade clay sculpture with fish and hole patterns, placed on a shelf. One of several sculptures P20 asked the AI to describe. Due to limitations in interpreting tactile or abstract objects, a sighted assistant was often needed to provide context. The image depicts a conversation between P20 and VisionPal. In this conversation, P20 is trying to understand the point of view of a sighted person. First, P20 asks “Does the sculpture have realistic features?” and then the AI responds “No, the sculpture has stylized fish-like features rather than realistic ones.” Second, P20 “can you describe me the parts one by one?” and then the AI responds “Certainly. The sculpture includes a a circular body, a fish head, fins, a tail, and inner details. Finally, P20 asks “Does it have scales?” and the AI responds “Yes, the inner details resemble stylized scales.”}
  \caption{After receiving a description for a captured photo, participants engaged with an MLLM. P20 often started conversations to follow-up with additional questions about his clay sculptures like "Can you describe me the parts one by one?" to understand how a sighted person would interpret their artwork.}
  \label{fig: sculptures}
  \Description{A Handmade clay sculpture with fish and hole patterns, placed on a shelf. One of several sculptures P20 asked the AI to describe. Due to limitations in interpreting how a sighted person would perceive this object, the user requested additional context from VisionPal to understand the sighted person perspective.}
\end{teaserfigure}

\maketitle

\section{Introduction} 

Blind and Low Vision (BLV) people encounter barriers in day-to-day activities because visual information is not readily available to them. The presence of these barriers within critical everyday tasks, like cooking meals or managing medications at home, \cite{kamikuboShopping, jeamwatthanachai2019indoor, kim2025toward, cooking2024Mingzhe} often undermines BLV people’s independence.

Over the past decade, AI-powered visual interpretation applications like Seeing AI \cite{seeingAI} have emerged to support BLV people’s access to visual information. Using these applications, users can identify objects and read product labels by receiving AI-generated captions for captured photos. However, past work shows these applications often generate inaccurate interpretations and miss important visual details \cite{gonzalez2024usecases,granquist2021evaluation}. In our prior diary study of how BLV people use visual interpretation applications \cite{gonzalez2024usecases}, we found that the accuracy of the models used in these applications was mediocre: across 316 diary entries, only 31.96\% (101) of the produced visual interpretations “matched how a human would describe a visual scene” (Accuracy 3 out of 3), with a mean accuracy of 1.96 out of 3. Similarly, \citet{granquist2021evaluation} found that Seeing AI and Orcam had inconsistent reading performance across visual scenarios, correctly identifying 100\% of words in printed material like magazines but only 13\% of words on curved surfaces like pill bottles. These limitations highlight inaccuracies and inconsistencies in current visual interpretation applications across real-life scenarios, undermining users' independence and trust in these technologies \cite{nair2023imageassist,lee2021image}.

Recent advances in computer vision with multimodal large language models (MLLMs) show promise in improving visual interpretation systems’ accuracy and usefulness. Prior benchmarking studies in image captioning have found that MLLMs, like GPT-4o and Gemini-2.0, generate better visual interpretations than previous traditional captioning models, often reaching similar quality to human-authored descriptions \cite{cheng2025caparena, bucciarelli2024personalizing}. Unlike previous captioning models, MLLMs can also process images alongside natural language instructions. This theoretically enables users to engage with systems in more interactive formats, such as back-and-forth conversations to request and receive goal-relevant details\cite{openai_gpt4v_system_card}. MLLMs have been integrated into mainstream visual interpretation applications (e.g., Seeing AI, Be My AI, Envision) \cite{be_my_ai_blog, applevis2025seeingai,envision_app_wayback_2025} to extend support for a greater variety of visually complex tasks (e.g., reading a pregnancy test result, helping post images online) \cite{sharma2025visual}.

Despite this promising technology, early interview studies with BLV people suggest that MLLMs still struggle to meet daily needs reliably \cite{sharma2025visual,xie2024emerging,xie2025beyond}. For example, \citet{xie2024emerging} interviewed BLV people who used MLLM-enabled systems and found that they encountered frequent overly verbose outputs, omissions of critical details, and hallucinations. These issues were prevalent for blurry, low-light photos, common characteristics of images taken by BLV users \cite{bigham2010vizwiz}. Similarly, exploratory studies by \citet{chang2025probing} found that real-time visual interpretation prototypes powered by GPT-4o handled simple static visual queries well, but failed in dynamic settings where users sought spatial guidance \cite{chang2025probing,chang2024worldscribe}. In addition to these technical shortcomings, the conversational nature of interactions with MLLMs can foster false confidence: users may interpret the specificity, length, and ``human-like'' style of responses as signals of trustworthiness \cite{longvqa_huh,cohn2024believing}. However, these same systems often provide vague responses, fabricate information, and exhibit ``sycophancy''— agreeing with users or telling them ``what they want to hear'' rather than providing accurate information \cite{sun2025friendly,chang2025probing, chen2025surfacing}. Altogether, we still lack a clear understanding of how BLV people engage with conversational MLLM-enabled visual interpretation systems in everyday contexts and how effectively these systems address their needs.

To address this gap, we pose the following research questions:

\begin{itemize}
\item \textit{\textbf{How well} are BLV people’s visual interpretation \textbf{needs addressed} by an MLLM-enabled visual interpretation application?}
\item \textit{\textbf{What are} BLV people’s \textbf{real-world goals} when using an MLLM-enabled visual interpretation application and \textbf{where} do they \textbf{engage} with it?}
\item \textit{\textbf{Why} and \textbf{when} do BLV people \textbf{converse with MLLM-enabled visual interpretation} applications to address their needs?}
\end{itemize}

In order to answer these questions, we conducted a two-week diary study and follow-up interviews with 20 BLV participants. During the study, participants used a visual interpretation application we developed. This application was designed to replicate the functionality of mainstream MLLM-enabled visual interpretation applications commonly used by BLV people (e.g., Seeing AI and Be My AI). After each use, participants were encouraged to complete a brief survey that captured the context of use. Each diary entry included the survey response, the user-submitted photo, the AI-generated description, and the conversation history between the user and the application. In total, we collected 551 diary entries. In our analysis, for each diary entry, we identified participants’ goals for using the application, the location where they used it, the accuracy of the AI-generated description, the types of questions they asked during the conversation (identification, reading, etc.), and the correctness of the responses of the MLLM to users questions.

We found that participants were “somewhat satisfied” with the information generated by the AI and found it “trustworthy,” with mean satisfaction and trust ratings of 4.13 out of 5 (SD=1.07) and 3.76 out of 5 (SD=0.96), respectively.  Examples of novel use cases included using the application for keeping watch of a guide dog while exercising (P17) and supporting various tasks while preparing for a move (P15). 

Participants started follow-up conversations with the AI in 68\% of diary entries (375 out of 551). They used these dialogues in high-stakes situations like for identifying medications, confirming dosages, and checking food safety (e.g., verifying allergen information or judging whether food was done). Follow-ups typically occurred when participants sought non-salient visual details or they wanted some label or text to be read verbatim. In contrast, participants were less likely to open a conversation in public due to privacy concerns with speech input, or when they only needed a quick overview of their surroundings.
 
While participants’ trust and satisfaction were high, the AI's accuracy varied considerably across scenarios. For instance, descriptions of participants’ photos were highly accurate, with a mean accuracy of 2.9 out of 3 (SD = 0.34) but 22\% of follow-up questions contained hallucinations. In particular, text and graphics interpretation was error-prone with 34.6\% of follow-up responses containing one or more hallucinations; these errors were subtle and hard to notice—like an incorrect address, incorrect cooking times for recipes, or incorrect medication dosages (e.g., 100mg vs 200mg). The system also behaved inconsistently around sensitive content across participants: for some, it abstained when asked to identify addresses or names in correspondence, while for others it disclosed sensitive details verbatim, including salary information or medical labels containing participants’ full names.

Our results demonstrate that our application, which used GPT-4o, exhibited high accuracy for descriptive tasks and was rated higher in user satisfaction and trust compared to the pre-LLM captioning model we used in our previous study \cite{gonzalez2024usecases}. However, our findings also revealed gaps in how models like GPT-4o behave during follow-up conversations: while these models excel at producing image descriptions, they struggle when users engage in  back-and-forth visual assistance requests, often required for more complex tasks. Based on these findings, we propose nine behaviors to produce actionable and safe guidance that MLLMs should exhibit when assisting BLV users. Together, these nine behaviors represent what we define as the "visual assistant" skill. To ensure these behaviors emerge in real-world visual interpretation systems during visual assistance tasks, we identify three key moments where model trainers, application developers, and users can intervene in the visual interpretation pipeline: 

\begin{itemize}
\item \textbf{Training models to learn behaviors}: incorporating training examples that model the nine proposed visual assistant behaviors and also incorporating examples from successful interactions between real BLV people and human visual assistants (golden standard).
\item \textbf{Instructing visual interpretation systems for context and values alignment}: incorporating visual assistant behaviors descriptions and relevant examples in system prompts together with user’s context to ensure pre-trained models are re-aligned to provide appropriate visual assistance.
\item \textbf{Designing user-facing applications with behavior controls, and built-in personalization}: allowing users to leverage developer-provided settings and interfaces to tune the visual interpretation systems’ behavior to meet their personal needs and preferences (such as matching desired behaviors to specific task types).
\end{itemize}

Our work offers a real-world foundation for understanding how MLLM-enabled visual interpretation applications are used, where they succeed, where challenges remain, and actionable directions for developing MLLM-powered visual assistants for BLV people.

\section{Contributions, and Limitations} 

This paper makes the following contributions:

\begin{itemize}
\item \textbf{Empirical insights into how Blind and Low Vision people use an MLLM-enabled visual interpretation application}.  Through a two-week diary study and follow-up interviews, we characterized 20 BLV participants’ usage patterns, goals, and conversations across diverse everyday contexts.
\item \textbf{Evidence of MLLM strengths and limitations in real-world visual interpretation}. We document high user satisfaction and trust alongside a critical accuracy gap: while MLLMs excel at image descriptions (mean=2.9/3), 22\% of conversational follow-up responses contained hallucinations.
\item \textbf{Design recommendations for future MLLM-enabled visual interpretation systems}. We define the "visual assistant" skill to guide future MLLM development for assisting BLV people access visual information and propose behaviors that MLLMs should exhibit when supporting BLV users.
\end{itemize}

We also acknowledge limitations of our work. While participants were encouraged to use the application to meet daily life needs, and even used the application in sensitive contexts, the participants’ perception of being surveilled (e.g., observer effect) and knowing their data was being recorded may have influenced how participants ultimately interacted with the system and for what contexts.  Second, while we recruited a diverse set of participants across age and gender, all of our participants were expert users of visual interpretation systems, missing a critical perspective from beginners who have never engaged with visual interpretation systems (e.g., lack of awareness or access barriers). Additionally, some of our participants submitted a lower number of entries compared to others (min=5, max=62, mean=22.05). Finally, our data collection period happened during the months of October through December. As these months constitute a busy holiday season in the USA, some participants were traveling during the diary study—leading to possibly atypical use cases for our tool which may not be as prevalent outside of this time frame.

\section{Related Work}
We contribute to the broader research on visual interpretation for BLV people, focusing on how technology enables BLV people to address everyday visual challenges. Our work builds on prior investigations of remote sighted assistance, as well as studies exploring AI-powered visual interpretation systems for BLV users.

\subsection{Human-Powered Visual Interpretation Systems}
BLV people access visual information in their daily lives is limited. To overcome these challenges, BLV people often turn to visual interpretation support, whether from sighted friends or remote assistants who act as their “eyes.” This support is typically provided through in-person visual guidance or via computer or mobile devices connected to a remote sighted assistant \cite{garaj2003system, bigham2010vizwiz}.

Early visual interpretation systems focused primarily on human-assisted navigation, with less attention allocated to broader interpretation tasks \cite{garaj2003system, hunaiti2006remote, bujacz2008remote, bujacz2008remotemobility, chaudary2016tele}. For instance, \citet{garaj2003system} introduced a system where BLV users shared their location with a remote sighted assistant for outdoor guidance, and used chest-mounted cameras indoors when GPS was unavailable. \citet{bujacz2008remote} conducted user studies comparing in-person, unguided, and remote-sighted assistance, confirming the value of remote support for navigation \cite{bujacz2008remote, bujacz2008remotemobility}. Despite in-person assistance being faster and less error-prone, BLV users’ orientation and environmental awareness was higher when receiving remote sighted assistance\cite{bujacz2008remote}.

Researchers later broadened visual interpretation systems beyond navigation, developing applications to support with everyday tasks. A key example is VizWiz. This system connected BLV users to crowdsourced sighted volunteers via mobile devices for on-demand visual assistance \cite{bigham2010vizwiz}. In practice, this assistance was often asynchronous—not by design preference, but because volunteers frequently responded after a delay—an unavoidable by-product that shaped many later system. This line of work spurred research into crowdsourcing for visual interpretation \cite{lasecki2013answering, brady2013visual, brady2014friendsourcing, brady2013investigating, brady2015using}, using volunteers to assist with everyday tasks \cite{bigham2010vizwizlocate, guo2016vizlens, gleason2016vizmap, burton2012crowdsourcing}. While helpful, these systems were limited by response delays and single-photo, single-question formats, making them less effective for visually complex or time-sensitive situations.

To overcome the limitations of asynchronous assistance, real-time systems have emerged as a more responsive alternative, enabling dynamic exchanges to better support BLV users’ goals. Applications like Be My Eyes and AIRA connect users with sighted assistants via live video \cite{bemyeyes, aira}. Researchers have begun examining when, why, and how BLV people use these systems \cite{nguyen2019large, lee2018conversations, lee2020emerging}. In a large-scale study of 10,022 AIRA calls, \citet{nguyen2019large} found that users relied on AIRA for a variety of daily tasks—including reading, attending social events, and shopping.

Further research has explored how BLV users and sighted assistants can better communicate and share personal context. \citet{lee2018conversations} found that remote sighted assistants (RSA) often prompt users for context, while users provide real-time feedback and hand gestures to support navigation. To improve these interactions, \citet{xie-helping-helpers} proposed tools like interactive 3D maps for RSAs, later exploring multi-assistant collaboration \cite{xie-pair-volunteers}. Recent work has also examined how AI can support human assistants by helping identify uncommon scenes and objects, and deliver more context-aware guidance \cite{kamikubo2020support, xie2022iterative, lee2022opportunities, yu2024human}. While human-powered systems offer personalized and flexible support, they face challenges with availability, training, and privacy.

\subsection{AI-powered Visual Interpretation Systems} 
AI-powered visual interpretation systems have played a central role in expanding access to visual information for BLV people, enabling greater independence across a range of everyday tasks. Applications like Seeing AI \cite{seeingAI} and TapTapSee \cite{TapTapSee2025} were among the first tools to provide BLV users with on-demand access to AI-powered visual interpretations, such as object recognition and text reading. These applications introduced a new interaction paradigm centered on the needs of BLV users, enabling them to capture visual content and access various types of visual information within a single application independently. However, evaluations of these kinds of systems have surfaced notable limitations \cite{gonzalez2024usecases, hong2024understanding, kupferstein2020understanding}. In a previous study, we deployed an application that leveraged Microsoft's latest captioning models to emulate features of Seeing AI and evaluated it with BLV users. We found participants were frequently skeptical (2.43 on a 4-point trust scale, SD=1.16) and “somewhat unsatisfied” (2.76 on a 5-point satisfaction scale, SD=1.49) with AI-powered visual interpretations due to frequent errors in the descriptions produced by the captioning model \cite{gonzalez2024usecases}. Similarly, \citet{kupferstein2020understanding} observed low levels of satisfaction and trust among BLV users, highlighting persistent shortcomings in pre-LLM computer vision-based description systems.

More recently, advancements in MLLMs and large vision-language models, such as GPT-4V \cite{openai_gpt4v_system_card}, have been incorporated into applications like Seeing AI and Be My AI, enhancing their visual interpretation capabilities \cite{applevis2025seeingai, be_my_ai_blog}. Unlike traditional computer vision approaches, vision-language models integrate image content with natural language, enabling richer, and interactive VQA \cite{carolan2024review}.

Through interview and user studies, researchers have begun to explore how these MLLM-enabled applications could impact BLV users’ day-to-day experiences. These studies revealed that such systems can effectively support a range of tasks, including accessing cooking instructions and reading food packaging \cite{li2024recipe}, locating personal objects and navigating scenes \cite{mathis2025lifeinsight, chang2024worldscribe}, and managing complex content creation workflows \cite{UseAi_das}. \citet{xie2024emerging} found that long-term users of Be My AI incorporated it into activities such as fashion coordination, parenting, and social media engagement; yet, they frequently encountered verbose outputs, vague descriptions, and difficulties interpreting physical interfaces, such as rotary controls in household appliances. \citet{contesting_alharbi} further identified how BLV users adopted verification strategies to manage AI mistakes, including leveraging non-visual senses or involving sighted assistance. Across these studies, participants described adapting their approaches to compensate for system limitations, whether by refining prompts, switching modalities, or seeking external help. While this body of work offers valuable insights into emerging practices and challenges, it remains largely grounded on retrospective interviews or short-term prototype evaluations.

MLLM-enabled visual interpretation systems are changing BLV people's access to visual information but prior research has largely relied on interviews that do not fully reflect real-world use. We lack an understanding of how effective these systems are in BLV people's lives. Our work addresses this gap by analyzing users' real-world interactions with an MLLM-enabled visual interpretation application, showing both the application's real-world performance and the everyday impact of MLLMs on visual interpretation systems for BLV people.

\section{Method}
\label{sec:methods}

We conducted a two-week diary study and follow-up interviews with 20 BLV participants to investigate how well MLLM-enabled visual interpretation applications address their visual interpretation needs, examine their real-world use cases for these applications, and understand when and why they engage in conversational interactions with such applications. To collect diary entries, we developed VisionPal, a MLLM-enabled visual interpretation application that allowed participants to submit a short survey about their experience with every use. Participants used VisionPal during a two-week period to receive visual interpretations and ask follow-up questions in a chat interface. We collected and analyzed 551 diary entries and 626 questions from 375 conversations with an MLLM.

\aptLtoX{\begin{table*}[ht]
    \caption[]{Participant demographics. The description of vision is self-reported by participants. }
    \centering    
    \begin{tabular}{|c|c|c|p{7cm}|p{6cm}|}
    \hline
        PID & Age & Gender & Description of Vision   & Visual Interpretation Systems \\ \hline
        P1  & 45& Female&  Blind/Light Perception Only; Partially Sighted& BeMyEyes/BeMyAI; Seeing AI\\ \hline
        P2  & 45& Female& Blind/No Light Perception& BeMyEyes/BeMyAI; Seeing AI; AIRA; Envision AI\\ \hline
       P3 & 33& Female& Blind/No Light Perception& BeMyEyes/BeMyAI; Seeing AI; AIRA\\ \hline
        P4 & 51& Male& Low Vision& Seeing AI\\ \hline
        P5 & 27& Male& Low Vision; Partially Sighted; Legally Blind; Color Blindness& BeMyEyes/BeMyAI\\ \hline
        P6 & 75& Female& Blind/No Light Perception& BeMyEyes/BeMyAI; Seeing AI; AIRA\\ \hline
 P7& 41& Male& 
 Blind/Light Perception Only; Low Vision; Partially Sighted; Legally Blind; Peripheral Vision Loss; Central Vision Loss&BeMyEyes/BeMyAI; Seeing AI; AIRA; Envision AI\\\hline
 P8& 32& Female& Blind/No Light Perception&BeMyEyes/BeMyAI; Seeing AI; Envision AI; Speakaboo\\\hline
 P9& 46& Female& Low Vision&BeMyEyes/BeMyAI; Seeing AI\\\hline
 P10& 19& Female& Legally Blind; Color Blindness; Night Blindness; Peripheral Vision Loss; Temporary Vision Impairment&BeMyEyes/BeMyAI; Seeing AI\\\hline
 P11& 53& Male& Low Vision; Partially Sighted; Legally Blind; Night Blindness; Peripheral Vision Loss; Central Vision Loss&BeMyEyes/BeMyAI; Seeing AI; AIRA\\\hline
 P12& 25& Female& Low Vision; Partially Sighted; Legally Blind; Night Blindness; Peripheral Vision Loss; Central Vision Loss&BeMyEyes/BeMyAI; Seeing AI; AIRA\\\hline
 P13& 50& Male& Low Vision; Legally Blind; Peripheral Vision Loss&Seeing AI\\\hline
 P14& 22& Male&  
 Blind/Light Perception Only&BeMyEyes/BeMyAI; Seeing AI\\\hline
 P15& 58& Male&  
 Blind/Light Perception Only; Legally Blind&BeMyEyes/BeMyAI; Seeing AI; AIRA\\\hline
 P16& 24& Female& Legally Blind; Color Blindness; Night Blindness; Peripheral Vision Loss&BeMyEyes/BeMyAI; Seeing AI; AIRA; Envision AI\\\hline
 P17& 32& Female& 
 Blind/Light Perception Only&BeMyEyes/BeMyAI; Seeing AI; AIRA\\\hline
 P18& 72& Female& Blind/No Light Perception&BeMyEyes/BeMyAI; Seeing AI; AIRA\\\hline
 P19& 35& Male& 
 Blind/Light Perception Only&BeMyEyes/BeMyAI; Seeing AI; AIRA; Speakaboo\\\hline
 P20& 44& Male& Blind/No Light Perception&BeMyEyes/BeMyAI; Seeing AI\\\hline
    \end{tabular}    
    \label{tab:participant-details}
\end{table*}
}{
\begin{table*}[ht]
    \caption[]{Participant demographics. The description of vision is self-reported by participants. }
    \centering    
    \begin{tabular}{|c|c|c|p{7cm}|p{6cm}|}
    \hline
        PID & Age & Gender & \parbox[c]{6cm}{\centering Description of Vision}   & Visual Interpretation Systems \\ \hline
        P1  & 45& Female& \parbox[c]{6cm} \centering Blind/Light Perception Only; Partially Sighted& BeMyEyes/BeMyAI; Seeing AI\\ \hline
        P2  & 45& Female& \parbox[c]{6cm} \centering  Blind/No Light Perception& BeMyEyes/BeMyAI; Seeing AI; AIRA; Envision AI\\ \hline
       P3 & 33& Female& \parbox[c]{6cm} \centering  Blind/No Light Perception& BeMyEyes/BeMyAI; Seeing AI; AIRA\\ \hline
        P4 & 51& Male& \parbox[c]{6cm} \centering  Low Vision& Seeing AI\\ \hline
        P5 & 27& Male& \parbox[c]{6cm} \centering  Low Vision; Partially Sighted; Legally Blind; Color Blindness& BeMyEyes/BeMyAI\\ \hline
        P6 & 75& Female& \parbox[c]{6cm} \centering  Blind/No Light Perception& BeMyEyes/BeMyAI; Seeing AI; AIRA\\ \hline
 P7& 41& Male& \parbox[c]{6cm} \centering 
 Blind/Light Perception Only; Low Vision; Partially Sighted; Legally Blind; Peripheral Vision Loss; Central Vision Loss&BeMyEyes/BeMyAI; Seeing AI; AIRA; Envision AI\\\hline
 P8& 32& Female& \parbox[c]{6cm} \centering  Blind/No Light Perception&BeMyEyes/BeMyAI; Seeing AI; Envision AI; Speakaboo\\\hline
 P9& 46& Female& \parbox[c]{6cm} \centering  Low Vision&BeMyEyes/BeMyAI; Seeing AI\\\hline
 P10& 19& Female& \parbox[c]{6cm} \centering  Legally Blind; Color Blindness; Night Blindness; Peripheral Vision Loss; Temporary Vision Impairment&BeMyEyes/BeMyAI; Seeing AI\\\hline
 P11& 53& Male& \parbox[c]{6cm} \centering  Low Vision; Partially Sighted; Legally Blind; Night Blindness; Peripheral Vision Loss; Central Vision Loss&BeMyEyes/BeMyAI; Seeing AI; AIRA\\\hline
 P12& 25& Female& \parbox[c]{6cm} \centering  Low Vision; Partially Sighted; Legally Blind; Night Blindness; Peripheral Vision Loss; Central Vision Loss&BeMyEyes/BeMyAI; Seeing AI; AIRA\\\hline
 P13& 50& Male& \parbox[c]{6cm} \centering  Low Vision; Legally Blind; Peripheral Vision Loss&Seeing AI\\\hline
 P14& 22& Male& \parbox[c]{6cm} \centering 
 Blind/Light Perception Only&BeMyEyes/BeMyAI; Seeing AI\\\hline
 P15& 58& Male& \parbox[c]{6cm} \centering 
 Blind/Light Perception Only; Legally Blind&BeMyEyes/BeMyAI; Seeing AI; AIRA\\\hline
 P16& 24& Female& \parbox[c]{6cm} \centering  Legally Blind; Color Blindness; Night Blindness; Peripheral Vision Loss&BeMyEyes/BeMyAI; Seeing AI; AIRA; Envision AI\\\hline
 P17& 32& Female& \parbox[c]{6cm} \centering 
 Blind/Light Perception Only&BeMyEyes/BeMyAI; Seeing AI; AIRA\\\hline
 P18& 72& Female& \parbox[c]{6cm} \centering  Blind/No Light Perception&BeMyEyes/BeMyAI; Seeing AI; AIRA\\\hline
 P19& 35& Male& \parbox[c]{6cm} \centering 
 Blind/Light Perception Only&BeMyEyes/BeMyAI; Seeing AI; AIRA; Speakaboo\\\hline
 P20& 44& Male& \parbox[c]{6cm} \centering  Blind/No Light Perception&BeMyEyes/BeMyAI; Seeing AI\\\hline
    \end{tabular}    
    \label{tab:participant-details}
\end{table*}}

\subsection{Participants and Recruitment}

We partnered with a nonprofit organization, the LightHouse for the Blind and Visually Impaired in San Francisco, to recruit our participants. Prior to the initial interview, participants submitted an eligibility form, where we collected participants' demographic information, including the visual interpretation technologies they use, and their visual condition. Participants also received a consent form detailing data privacy protections (e.g., diary entry data, interview transcripts, etc.). Our eligibility criteria required participants to be 18 years or older, self-identify as Blind or Low Vision, have prior experience using Seeing AI or Be My AI, and own an iOS device running version 15 or higher to install and use our application. Participants received \$100 worth of compensation for their participation. This study was approved by Cornell University’s Institutional Review Board (IRB).

We recruited 20 BLV participants (eleven female, nine male), and participants were 19 to 75 years old (mean= 41.5, SD= 15.2). Thirteen participants identified as Blind, and 7 identified as having Low Vision.  One participant was located in Canada, one in Turkey, and the rest of the participants were located in the USA. See Table \ref{tab:participant-details} for more details.

\subsection{Procedure}
\begin{figure*}
\includegraphics[width=350pt]{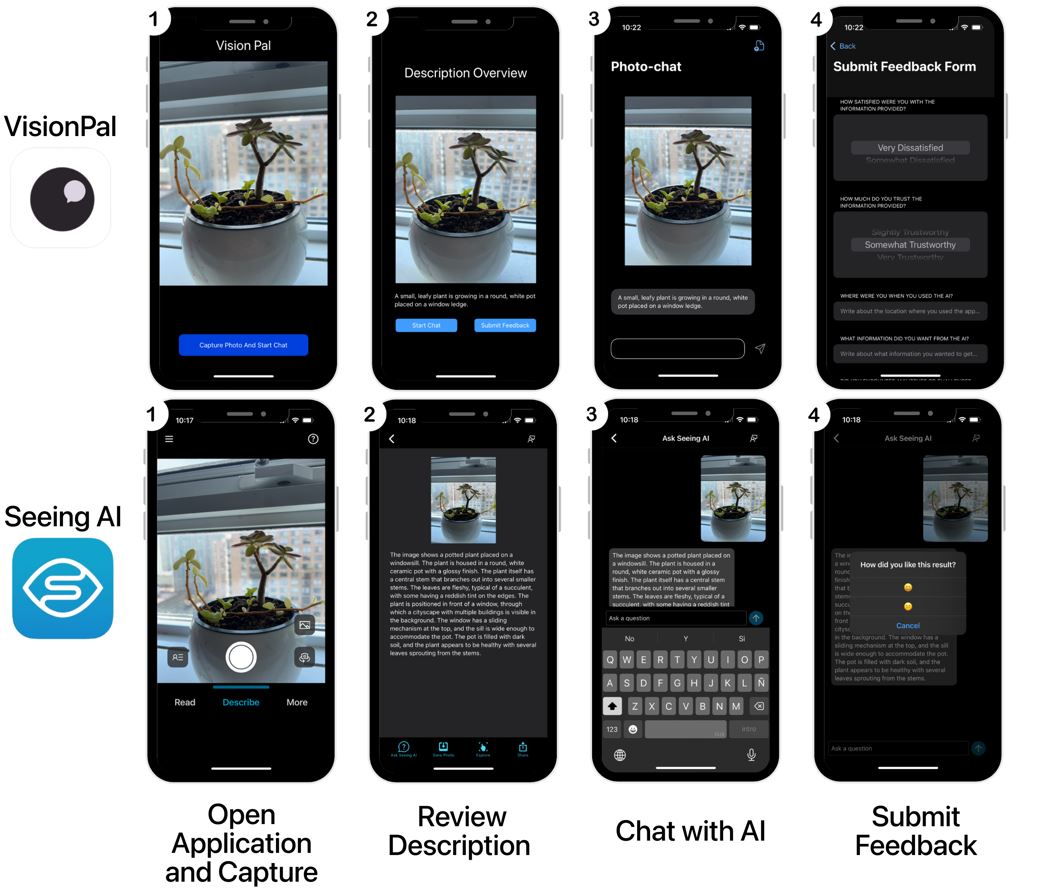}
\Description{8 phone screens, 4 side by side stacked. On top, visionpal. At the bottom, SeeingAI. Figure demonstrates similarities of both apps. Our application, VisionPal, is designed based on SeeingAI. In both applications, (1) the user opens the application to begin a visual assistance task (camera view can be seen on both), (2) they take a photo of their surroundings, and the app provides an overview of key elements in the image (now both screens show a description of an entity captured, seeingai has a longer description), (3) the user chats with the MLLM to ask follow-up questions or clarify specific details (Both interfaces are almost identical and look like a regular chatting app like iMessage), and (4) finally, they provide feedback on the AI’s performance, helping improve future interactions (SeeingAI shows a simple smiley face feedback while Visionpal shows a form that needs to be filled-out).}
\caption[]{\label{fig: interface} Our application, VisionPal, interaction design is based on Seeing AI. In both applications, the user (1) opens the application to begin a visual assistance task, (2) takes a photo of their surroundings to receive an overview of key elements in the image, (3) chats with an MLLM to ask follow-up questions or clarify specific details, and (4) provides feedback on on their experience to improve future interactions.}
\end{figure*}

In this diary study, which followed the procedure from our prior work \cite{gonzalez2024usecases}, participants completed an introductory session, a two-week diary period, and a post-diary interview. Data collection ran from October to December 2024.

During the introductory session, participants were given access to VisionPal, our data collection application. We conducted remote sessions through Zoom, lasting approximately 60 minutes. During the session, we asked participants to elaborate on their visual condition to understand their level of visual impairment (e.g., “Can you read with magnification?”). We also probed about participants' perception of MLLMs for visual interpretation and their impact in their personal lives (e.g., “Have MLLMs had an impact on your daily life?”). Next, we did a walkthrough of the interface of VisionPal and instructed participants how to interact with it. We finished the session after participants submitted a diary entry on their own.

During the two-week diary period, participants were recommended to complete at least one survey per day. In the introductory session, they were given the following guidance to encourage naturalistic use while still remaining safe: “\textit{use the application as much as you feel comfortable to meet your needs (...) but do not take any critical decisions based solely on the information provided by the AI}”. To help participants meet this daily usage goal, we programmed VisionPal to send daily notifications. Each diary entry consisted of the participant's captured photo, the photo description, the chat messages exchanged between the participant and the MLLM, and participants' responses to a short survey. We asked six questions, aiming to capture both the outcome of the interaction and the context of where and why participants engaged with the application:

\begin{itemize}
\item How satisfied were you with the information provided? \textit{5-point Likert scale (Very dissatisfied, somewhat dissatisfied, neither dissatisfied nor satisfied, somewhat satisfied, and very satisfied)}
\item How much do you trust the information provided? \textit{5-point Likert scale (Not at all trustworthy, slightly trustworthy, somewhat trustworthy, trustworthy, extremely trustworthy)}
\item Where were you when you used the AI? \textit{Open-ended question}
\item What information did you want from the AI? \textit{Open-ended question}
\item Did you encounter any issues or challenges during this interaction? \textit{Open-ended question}
\item Add any additional feedback. \textit{Open-ended question}
\end{itemize}

To prepare for the follow-up interviews, one researcher reviewed each participant’s diary entries and flagged salient interactions and feedback to guide interview probes. Interviews were conducted remotely via Zoom and lasted 60 to 90 minutes. To help participants recall their experiences, we enabled an interface in VisionPal that allowed them to revisit their submitted diary entries during the interview. These interviews focused on participants’ experiences using VisionPal throughout the two-week period, including their motivations and actions while interacting with VisionPal.

We divided the interviews into two sections. First, participants reflected on their overall experience with the application, discussing what they liked and disliked about their interactions and sharing memorable use cases (e.g., failures, successes, and surprises). Then, we explored participants' diary entries in depth, with the interviewer using prepared notes and questions to probe for additional context from selected entries as needed.

\subsection{VisionPal: Our Data Collection Application}

To study how BLV people engage with MLLM-enabled visual interpretation applications in real-world contexts, we developed a custom application, \textit{VisionPal}, to collect in-situ usage data. Existing commercial applications (e.g., Be My AI) are not open source and do not provide mechanisms for researchers to collect usage data. To address these limitations, VisionPal was developed based on our prior work's approach for consistent in-situ data collection for visual interpretation application usage \cite{gonzalez2024usecases}.

VisionPal’s design aligned closely with the interaction flow of commercial applications, like Seeing AI (See Figure~\ref{fig: interface}), to make the experience of using VisionPal closely resemble tools currently used by BLV people in daily life. Upon launching VisionPal, participants can capture a photo and receive a visual description generated by the MLLM. Following this, users can engage in a follow-up conversation with the model or proceed to fill-out a short survey (Figure~\ref{fig: interface}). The conversational interface allowed participants to exchange messages with the MLLM. Upon concluding the interaction, participants completed a six-question survey and submitted their diary entries. Diary entries data stayed on-device and were also sent encrypted to cloud storage anonymized. 

To power the visual interpretations of VisionPal, we chose the GPT-4o model. In past studies, GPT-4v, its predecessor, has demonstrated robust performance across a range of vision-language tasks \cite{longvqa_huh, openai_gpt4v_system_card}, including prototype implementations for real-time visual interpretation tasks in systems such as WorldScribe \cite{chang2024worldscribe} and LifeInSight \cite{mathis2025lifeinsight}. GPT-4v supports a wide range of visual tasks, including optical character recognition, abstract visual reasoning, and scene understanding \cite{yang2023dawn}. As commercial applications like Seeing AI and Be My AI use models like GPT-4v  \cite{be_my_ai_blog, openai_gpt4v_system_card, seeingaigpt4}, GPT-4o was an appropriate choice for approximating the capabilities of contemporary visual interpretation systems in a research context. In VisionPal we specifically used the GPT-4o model version “gpt-4o-2024-08-06” with a knowledge cutoff on September 30, 2023 \cite{openai_gpt4o}.

To generate photo descriptions with GPT-4o, the first API call contained the user's first image along with an user role message containing the following prompt:

\begin{quote}
\texttt{``Can you describe what is in this image? Limit your response to one sentence and don’t provide any information that might be obvious to a blind person. Provide information that would be very likely useful to a blind or low vision person.``}
\end{quote}

We designed this prompt to generate brief but useful descriptions—similar to WorldScribe’s approach, which limited object descriptions to one sentence \cite{chang2024worldscribe}. While verbose initial descriptions are generally more informative, we wanted to encourage participants to pose questions to the AI to study their information seeking needs. Based on pilot tests with two BLV individuals, this prompt generated descriptions with sufficient context for users to ask follow-up questions if needed. We provided no additional instructions or restrictions on model behavior. After receiving each photo description, participants could ask follow-up questions to the model through a chat feature. To continue generating relevant responses, we prepended the image and all the chat history to the API call for each subsequent message participants sent.

\subsection{Data}

Our 20 participants interacted with VisionPal 551 times during their diary study period. We reviewed all entries submitted from right after the initial session and to right before the follow-up interview, including incomplete entries where participants engaged with the application but did not submit the short survey. Our final dataset included 446 complete diary entries (mean=22.05, SD=11.54) and 105 incomplete entries (mean=5.50, SD=7.34). Complete entries per participant ranged from 4 (P1) to 45 (P20). See Table~\ref{tab:DiaryEntries} for details.

In addition to the entries, we analyzed transcripts from all 20 interviews to complement our understanding of participants' experiences and reveal deeper context behind their interactions.

\subsection{Analyzing Diary Entries Context}
\begin{table}
\caption{Diary entries submitted by participants. Incomplete entries typically include an image, a description, and a conversation history but may be missing participants' survey submission (feedback and responses to context questions).}
\label{tab:DiaryEntries}
\begin{tabular}{cccc}
\toprule
\textbf{PID} & \textbf{\# Diary entries} & \textbf{\# Incomplete entries} & \textbf{Total} \\
\midrule
1  & 4  & 1  & 5  \\
2  & 28 & 7  & 35 \\
3  & 8  & 0  & 8  \\
4  & 37 & 5  & 42 \\
5  & 7  & 1  & 8  \\
6  & 21 & 4  & 25 \\
7  & 14 & 0  & 14 \\
8  & 29 & 33 & 62 \\
9  & 23 & 10 & 33 \\
10 & 10 & 12 & 22 \\
11 & 28 & 8  & 36 \\
12 & 25 & 2  & 27 \\
13 & 32 & 1  & 33 \\
14 & 9  & 1  & 10 \\
15 & 41 & 6  & 47 \\
16 & 15 & 3  & 18 \\
17 & 24 & 2  & 26 \\
18 & 19 & 2  & 21 \\
19 & 22 & 8  & 30 \\
20 & 45 & 4  & 49 \\
\bottomrule
\end{tabular}
\end{table}

Our analysis built upon the methodology of our past diary study with BLV people on their use of scene description applications \cite{gonzalez2024usecases}, extending our approach to examine the conversational interactions enabled by VisionPal.

We coded the diary entries' context in three stages. First, three researchers collaboratively coded diaries from two participants to develop consistency on the coding practice. Second, the remaining diaries were divided among the researchers. Finally, one researcher reviewed all codes for consistency, and two researchers validated the revisions. In the following sections, we explain our coding criteria in detail.

\subsubsection{Coding Context of Use: User Goal and Location}
To code the context of diary entries, we initially applied closed codes from our previous study's codebook \cite{gonzalez2024usecases}, then created additional codes to categorize user goals and locations not captured in our prior work (e.g., reframed the category “Used in travel service sites” to “Transit settings”). For each completed diary entry, we identified the main user’s goal drawn from the participants’ response to the question: “What information did you want from the AI?” to understand how their motivation related to the follow-up questions they asked. Additionally, we identified the user's location through their self-reported response to the question “Where were you when you used the AI?” For entries where participants did not submit a survey or left the response fields empty, we coded the user’s goal as “Unknown goal” and location as “Unknown location”. In some cases, location was inferred based on previous or follow-up diary entries submitted within a short span of time. See Appendix \ref{appendix:Location Categories} and \ref{appendix:User Goal Categories} for full examples of locations and user goals.

\subsubsection{Categorizing User Questions}  
To categorize the kinds of requests users posed in their follow-up questions, we adapted \citet{chen2025fully}'s taxonomy for visual-question-answering types of user questions and conducted two  rounds of coding. In the first round, we applied some of their categories (Verification, Identification, Evidence-based, Instruction, and Advice).

Many questions initially categorized as "Evidence-based" reflected different visual tasks like spatial understanding, feature identification, and text recognition. To better evaluate the quality of the AI’s assistance to users, we conducted a second round of coding to introduce additional categories aligned with real-world visual interpretation needs. See Appendix \ref{appendix:User Question Categories} for our complete categories.

\subsection{Analyzing MLLM Visual Interpretations Accuracy}
\begin{figure}
\includegraphics[width=\linewidth]{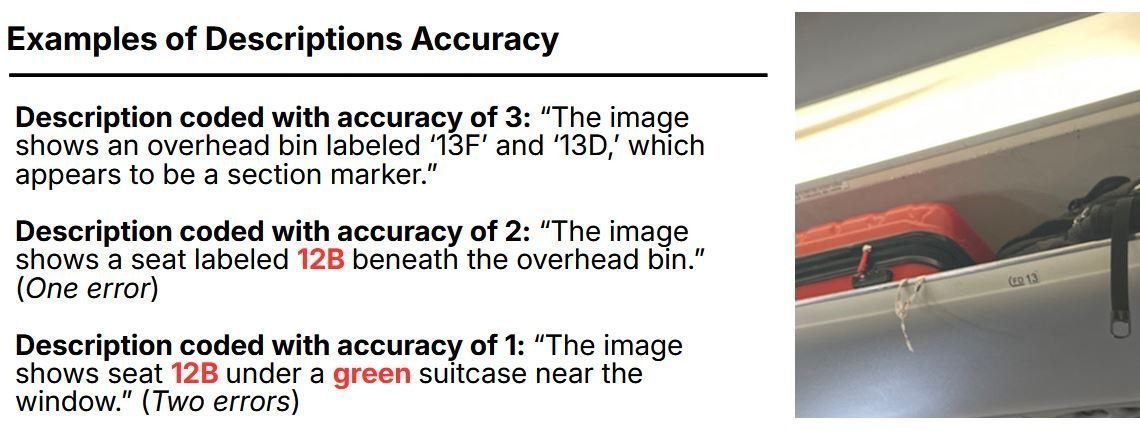}
\Description{ Coding approach to determine initial description accuracy. AI responses vary in accuracy when describing the image depicted to the right. We can see an overhead compartment open with a red suitcase and a dark green bag next to it. We can also see the bin with the sit label ‘13FD’. There are three example descriptions. Accuracy 3 description: “the image shows an overhead bin labeled ‘13f’ and ‘13d’, which appears to be a section marker. Accuracy 2 description: “The image shows a seat labeled 12B beneath the overhead bin”. For this description, only the seat label is inaccurate. Accuracy 1 description: “The image shows seat 12B under a green suitcase near the window”. For this description both the seat and the suitcase description are inaccurate. Note that accuracy is only with regards to correctness, not necessarily completeness}
\caption[]{\label{fig: accuracy ex} Demonstrative examples of our coding approach for initial description accuracy. Highlighted in red, the first photo description contains no hallucinations and accurately read the seating section information on the overhead bin compartment, the second one misidentifies the seat number (one hallucination), and the third description misidentifies the seat number and misidentifies the color of the suitcase (two hallucinations).}
\end{figure}

\begin{figure}
\includegraphics[width=\linewidth]{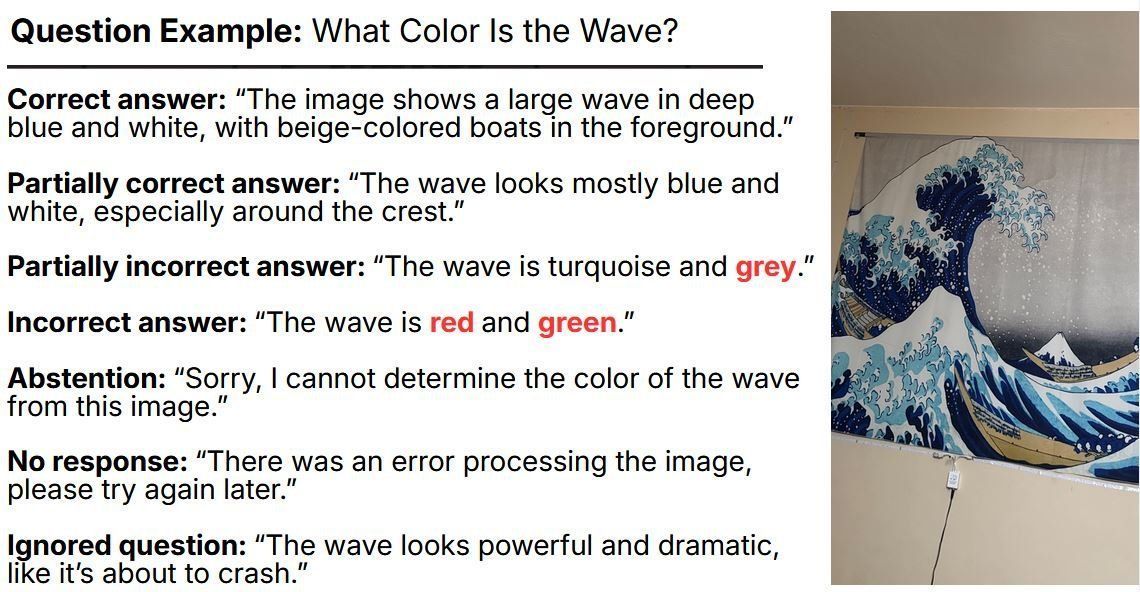}
\Description{Coding approach to determine responses correctness. Above is an example of a question with example responses with different levels of truthful information and completeness. The question is “what color is the wave?”. In the correct answer the wave is described with all colors correctly (deep blue, white, and beige accents). In the partially correct all colors are correct but some are missing (blue and white). In the partially incorrect, some are correct and some wrong (turqoise and grey). In the incorrect answer, all colors are wrong (red and green). In the abstained example, the ai responds: “Sorry, I cannot provide the color of the wave”. In the no answer there is a message error. In the ignored question the AI responds: “The wave looks powerful and dramatic”.}
\caption[]{\label{fig: correctness} Our coding approach to determine response correctness. We present an example question with example responses demonstrating varying levels of truthful and complete information. }
\end{figure}

To understand VisionPal's accuracy in visual interpretation tasks, we coded three key aspects of diary entries data: the accuracy of VisionPal’s photo descriptions, whether user questions could be answered from the available visual information, and the correctness of VisionPal's responses to answerable questions. For these aspects of visual interpretation accuracy, we followed the same three-stage coding process for developing a consistent coding criteria: three researchers coded two diaries together to ensure consistency, then remaining diary entries were split among the three researchers, and finally one researcher reviewed all codes and validated with the other researchers.

\subsubsection{Photo Description Accuracy}
To understand how well VisionPal generated a reliable baseline for the visual context captured, we coded the accuracy of photo descriptions using a modified version of \citet{gubbi2024context}'s methodology. To do this, we identified the number of hallucinations present in the description and assigned a score based on a 3-point scale: 3 (no hallucinations), 2 (one hallucination), and 1 (two or more hallucinations). We defined a hallucination as the AI including information without matching visual evidence \cite{gubbi2024context}. See Figure~\ref{fig: accuracy ex} for examples.

\subsubsection{Question Response Analysis}

\begin{figure}
\includegraphics[width=\linewidth]{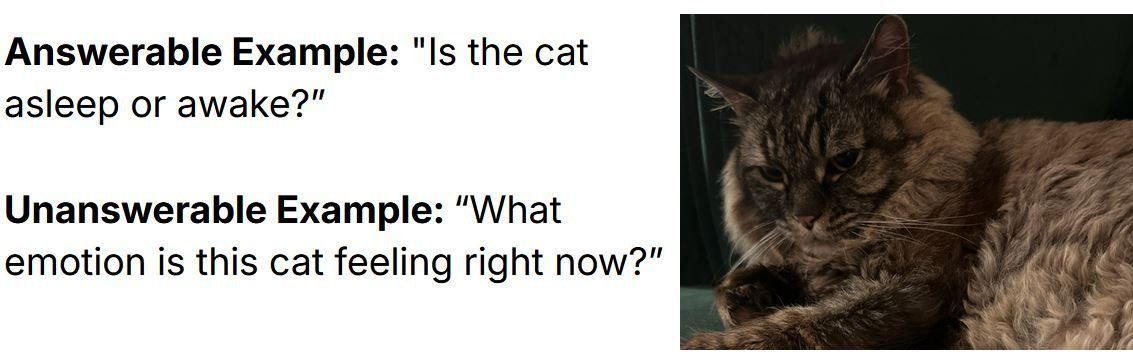}
\Description{The first question "Is the cat asleep?" is answerable because there is visual evidence that the cat is awake (their eyes are open). The second question "How is the cat feeling?" is unanswerable because there is no visual evidence of the emotions the cat is feeling.}
\caption[]{\label{fig: answerability} An example of our coding approach for determining question answerability. For the first question "Is the cat asleep or awake?", there is visual evidence that the cat is awake so the question is labeled as "answerable". For the second question "What emotion is this cat feeling right now", there is no visual evidence of the emotions the cat is feeling so the question is labeled as "unanswerable".}
\end{figure}

To code whether VisionPal answered questions correctly, we first determined if questions were answerable or unanswerable (see Figure~\ref{fig: answerability}). Questions were answerable if sufficient visual evidence existed to generate a response, and unanswerable if the visual context was too ambiguous to generate a definitive answer (e.g., overly blurry photos, close-ups, missing necessary visual context to answer).

For answerable questions, we coded response correctness using seven categories (see Figure~\ref{fig: correctness}): \begin{itemize} 
\item \textbf{Correct:} Fully addressed the question without errors. 
\item \textbf{Partially correct:} Missed some aspects but contained no false information (e.g., identifying some but not all colors of a shirt, reading some but not all menu categories). 
\item \textbf{Partially incorrect:} Contained both correct and incorrect information (e.g., cooking instructions with wrong temperatures or times). 
\item \textbf{Incorrect:} Only contained false information. 
\item \textbf{Abstained:} VisionPal explicitly stated it could not provide the requested information. 
\item \textbf{No response:} the API failed to respond or timed out.
\item \textbf{Ignored:} VisionPal failed to acknowledge the question and only provided unrelated information.
\end{itemize} 
This approach allowed us to capture the nuanced ways VisionPal handled user questions, distinguishing between technical failures, deliberate abstentions, and varying degrees of response accuracy.

\subsection{Analyzing Interviews}

To complement our diary entries analysis, we coded follow-up interviews using inductive open coding \cite{saldana2021coding}, followed by affinity diagramming \cite{hartson2012ux} to derive themes. Three researchers coded five interviews together to develop a shared codebook, then split the remaining interviews for individual coding. After all interviews were coded, the three researchers met to conduct two rounds of affinity diagramming to identify themes among participants' responses, and to understand why participants engaged the MLLM in follow-up questions.

\section{Findings}
We present findings on participants’ satisfaction and trust, their conversational interactions, and the types of questions they asked, illustrating how they engaged with our MLLM-enabled visual interpretation application and its effectiveness addressing user needs in real-world scenarios.

\subsection{Accuracy, Satisfaction, and Trust}

\subsubsection{VisionPal’s Accuracy Describing Photos}

\begin{figure}
\includegraphics[width=\linewidth]{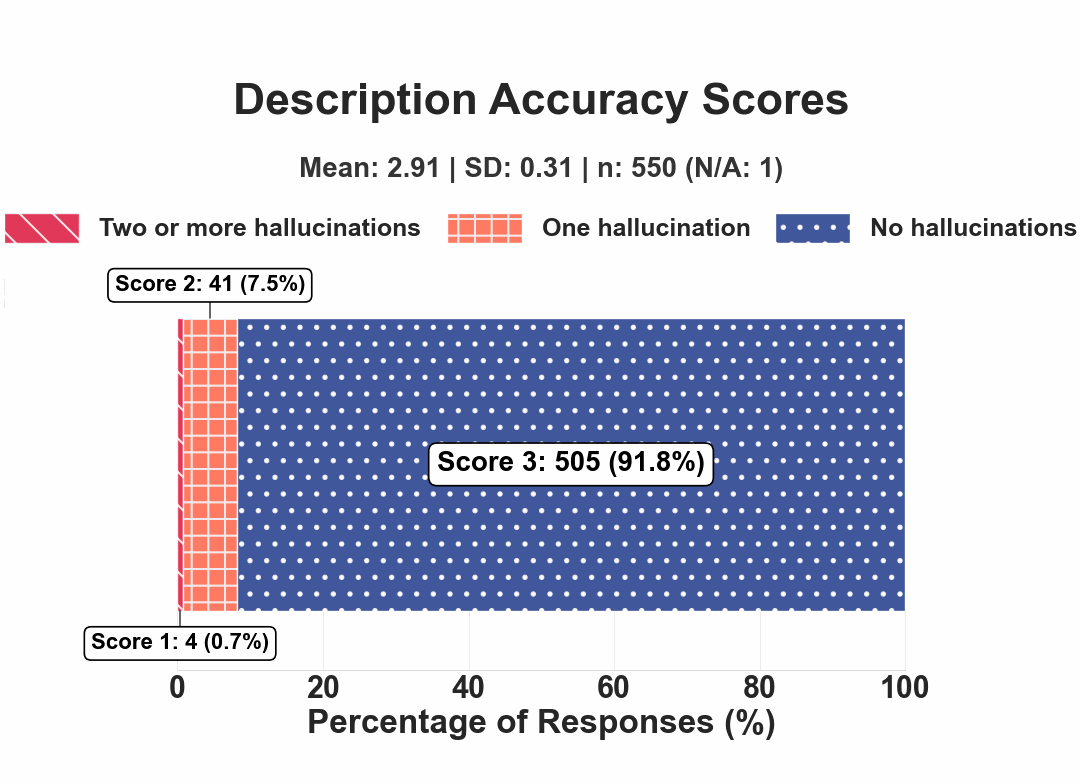}
\Description{Summary of initial description accuracy scores. Horizontal bar chart titled "Initial Description Accuracy Scores" showing accuracy ratings on a 3-point scale (1=lowest, 2=medium, 3=highest) categorized by hallucination presence (two or more hallucinations, one hallucination, no hallucinations). Results (n=550): Score 1: 4 descriptions (0.7\%), Score 2: 41 descriptions (7.5\%), Score 3: 505 descriptions (91.8\%). Mean=2.91, SD=0.31. Most descriptions were highly accurate, with 91.8\% receiving the maximum score of 3 (no hallucinations).}
\caption[]{\label{fig: Accuracy}  Summary of initial description accuracy scores. Most descriptions were highly accurate, with 91.8\% receiving the maximum score of 3 (no hallucinations).
}
\end{figure}

Participants frequently received accurate descriptions of the photos they submitted. The mean accuracy of initial descriptions was 2.91 out of 3 (SD = 0.31), and 505 entries (91.8\%) received a score of 3. Only 4 entries (0.7\%) received the lowest score of 1, typically reflecting descriptions that had multiple misidentifications (e.g., the AI “hallucinated” multiple inexistent objects, text, or visual features). For example, in one entry from P20, the AI misidentified a battery charger for watch batteries as a dial controller, describing it as a device for adjusting volume or light levels. The remaining 41 entries (7.5\%) were rated a 2, having only one hallucination present in the description. For photo descriptions, the application provided accurate information in most cases, showing the overall proficiency of MLLMs at describing visual content. See Figure \ref{fig: Accuracy} for details.

\subsubsection{Participants’ Satisfaction with MLLM-enabled Visual Interpretations}

\begin{figure}
\includegraphics[width=\linewidth]{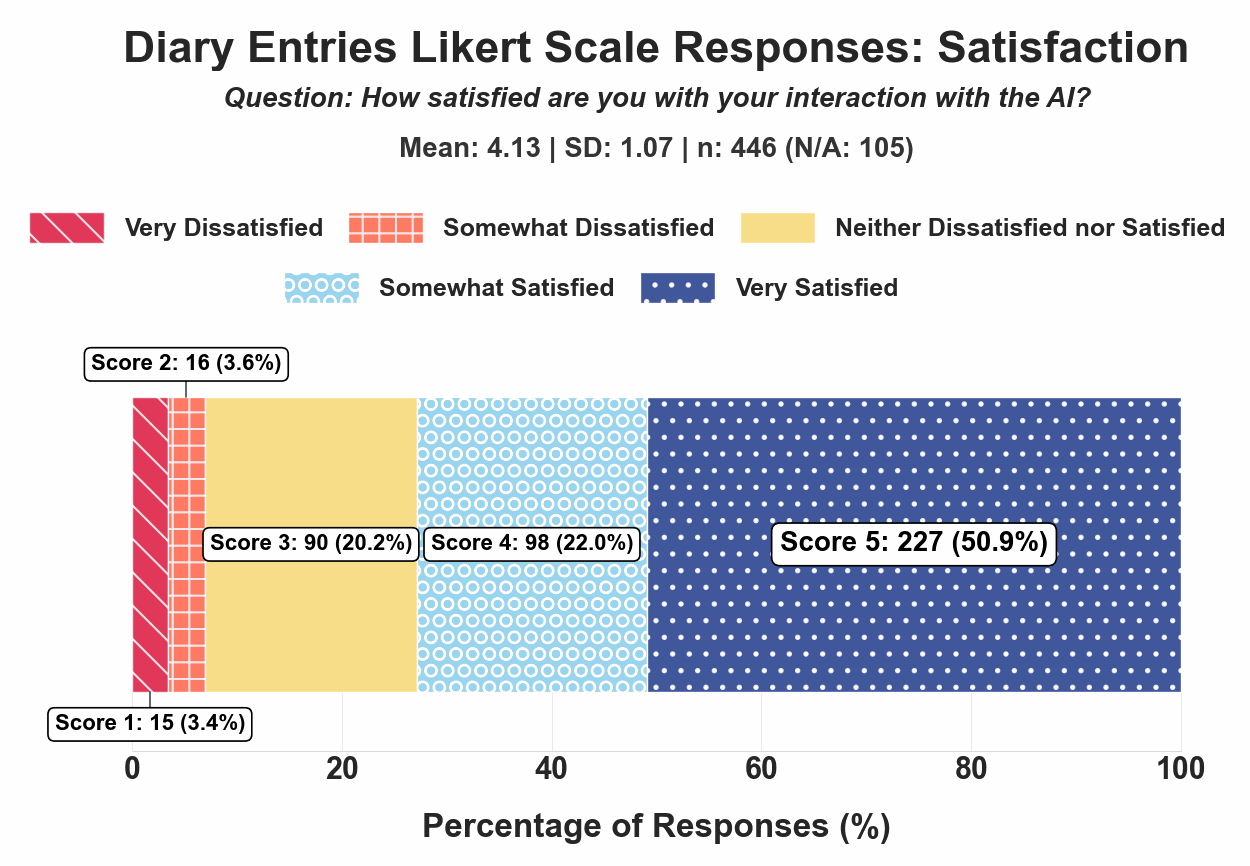}
\Description{Summary of participants’ satisfaction ratings. A horizontal bar chart titled "Diary Entries Likert Scale Responses: Satisfaction" for the question "How satisfied are you with your interaction with the AI?" 5-point scale from Very Dissatisfied to Very Satisfied. Results (n=446): Score 1 (Very Dissatisfied): 15 entries (3.4\%), Score 2 (Somewhat Dissatisfied): 16 entries (3.6\%), Score 3 (Neither): 90 entries  (20.2\%), Score 4 (Somewhat Satisfied): 98 entries (22.0\%), Score 5 (Very Satisfied): 227 entries (50.9\%). Mean=4.13, SD=1.07. Most ratings were positive or neutral, with over 50.9\% of completed entries receiving the maximum score of 5.}
\caption[]{\label{fig: Satisfaction}  Summary of participants’ satisfaction ratings. Most ratings were positive or neutral, with over 50.9\% of completed entries receiving the maximum score of 5. Low ratings were rare (< 7\% of entries) and typically linked to failures in accuracy where the users had some working knowledge of the correct answer.
}
\end{figure}

Participants submitted 446 complete diary entries and generally reported positive experiences with the AI. Satisfaction ratings averaged 4.13 out of 5 (SD = 1.07), indicating that most participants were somewhat or very satisfied using the application. Specifically, 227 diary entries received the highest satisfaction score of 5, while 98 entries were rated a 4, and 90 entries were rated a 3. In contrast, only a small number of entries were rated as 1 (15 entries) or 2 (16), suggesting low dissatisfaction overall. These lower scores were often associated with inaccurate answers that did not meet participants' expectations, and users were aware that the visual interpretation contained hallucinations. For instance, in one conversation (P5), the AI misidentified an airplane seat number, with the resulting inconvenience and confusion during the boarding process leading to a satisfaction score of 1. Other instances of lower scores were associated with interactions where users felt obligated to ask multiple follow-up questions to get a full overview of their visual surroundings or acquire the visual information they needed. P17 in their diary entry feedback said: \textit{“No issues. It just feels like pulling teeth to get all of the information!”} See Figure \ref{fig: Satisfaction} for details.

\subsubsection{Participants’ Trust in MLLM-enabled Visual Interpretations}

\begin{figure}
\includegraphics[width=\linewidth]{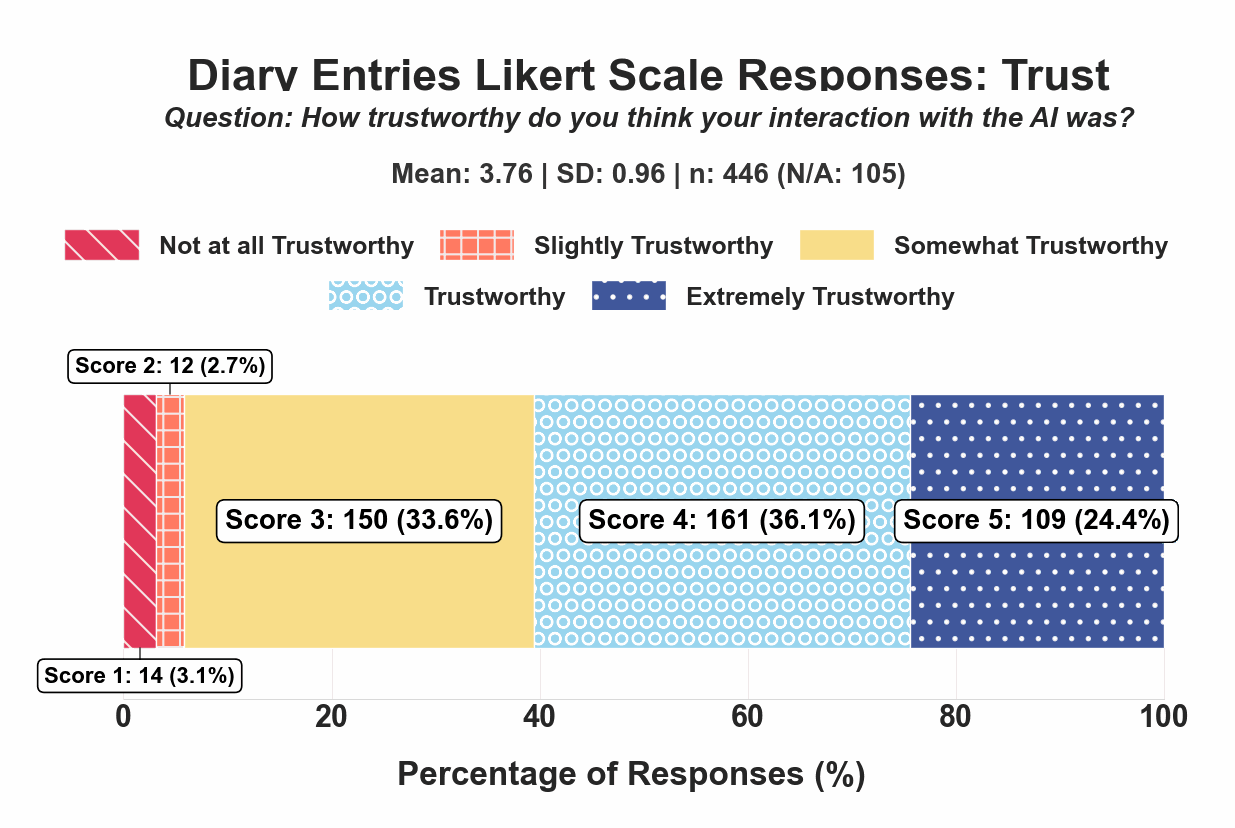}
\Description{ Distribution of  Trust by User Goals. 100\% stacked bars (not at all to extremely trustworthy) across the same goals. Trust peaks for “Learn information about subject” (≈67\% extremely, 33\% trustworthy) and “Describe a person” (50\% extremely, 25\% trustworthy); guidance and scenery also rate highly; interfaces are evenly split among somewhat/trustworthy/extremely; identification tasks lean positive; text reading, app-learning, and especially navigation show lower intensity (more “somewhat,” little or no “extremely”).}
\caption[]{\label{fig: Trust}  Summary of participants’ trust ratings. Most ratings were positive or neutral: 24.4\% of completed entries receiving the maximum score of 5. Although high ratings were less common than for satisfaction (signaling participants skepticism), low ratings were very uncommon (<7\% of entries).
}
\end{figure}

Trust ratings were also positive, though slightly lower than ratings for satisfaction. Participants rated the trustworthiness of AI visual interpretations at 3.76 out of 5 (SD = 0.96). Most ratings fell between “somewhat trustworthy” and “extremely trustworthy”: participants rated 150 entries a 3, 161 a 4,  and 109 a 5. Very few entries received low trust scores: only 14 entries were rated as “not trustworthy at all” (1)  and 12 as “a little trustworthy” (2). These lower trust ratings typically stemmed from responses that were inaccurate and users immediately verified the generated responses by acting on the information or asking a sighted peer.  For example, P10 gave a low trust score when the AI misidentified the location of a button on a car stereo, offering misleading guidance to change the volume of the music. Similarly, P11 wanted to check a cheese’s freshness by asking, \textit{“Is there any blue mold on the cheese?”}) The application replied, \textit{“No, there is no visible blue mold on the cheese,”} but a nearby friend confirmed the presence of mold. See Figure \ref{fig: Trust} for details.

In the following Section, we examine the context in which VisionPal was used, focusing on participants' goals and locations.

\subsection{MLLM Daily Life Use Context}
\subsubsection{User Goals}
\begin{figure*}
\includegraphics[width=0.9\linewidth]{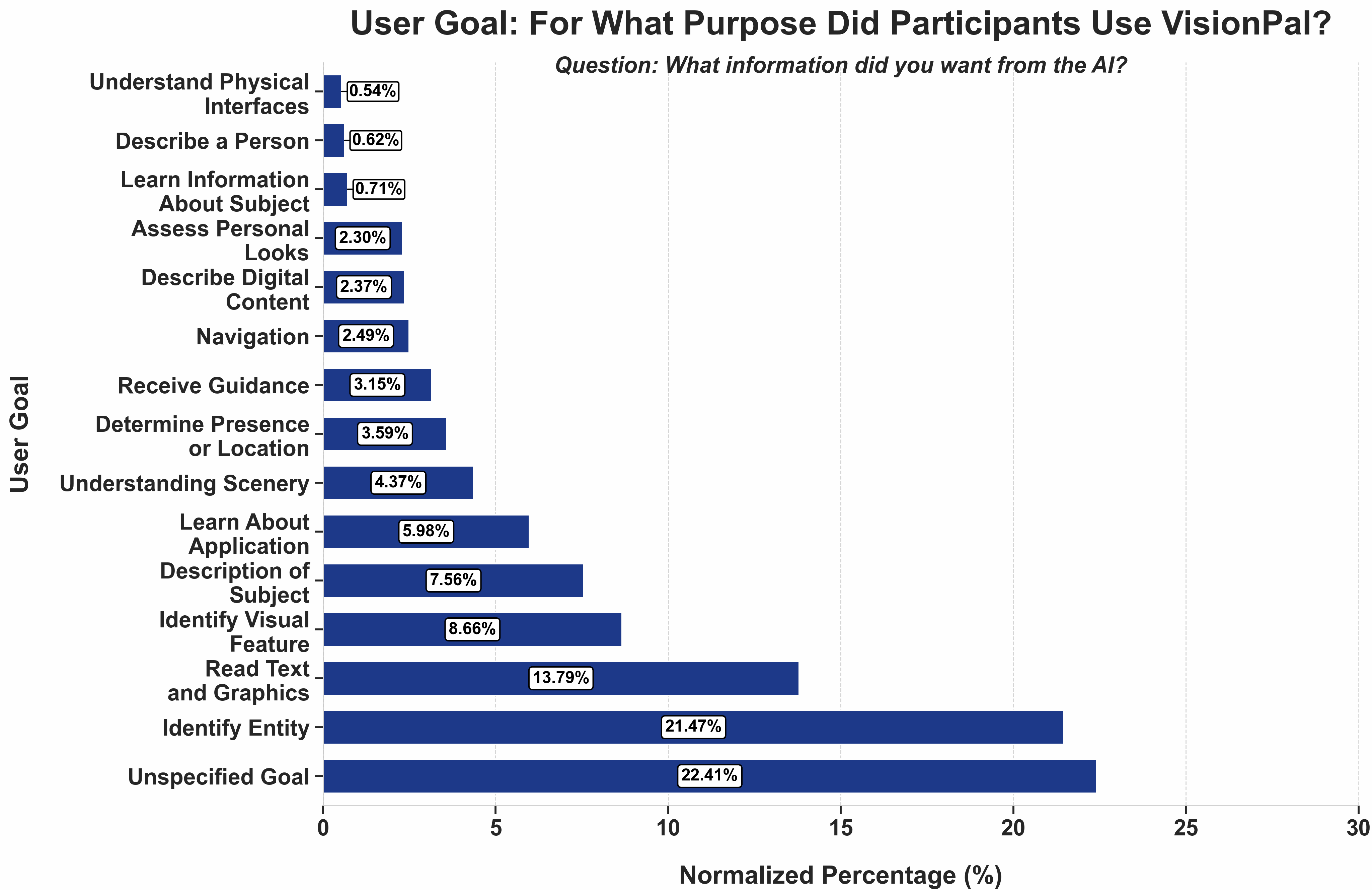}
\Description{The distribution of participants' goals when using the VisionPal application. The most common goals were identifying an entity by name and accessing information encoded in text, numbers, or other types of graphics. **The goal percentages were normalized to avoid skewing frequency towards participants that interacted with the application the most. Percentages were: Unknown goal (22.53\%), Identify entity (20.49\%), Read Text, numbers, and graphics (13.46\%), Identify feature 9.11\%, Description of subject 7.96\%, Learn application 6.30\%, Build understanding of scenery 4.6\%, receive guidance 3.31\%, determine presence and location 2.73\%, describe digital content 2.49\%, Navigation 2.62\%, personal appearance 2.42\%, describe person 0.66\%, Learn information about subject 0.75\%, and understand physical interface 0.57\%.}
\caption[]{\label{fig: Distribution of User Goals} The distribution of participants' goals when using VisionPal. The most common goals were identifying an entity by name and accessing information encoded in text, numbers, or other types of graphics. **The goal percentages were normalized to avoid skewing frequency towards power users.
}
\end{figure*}
\begin{figure*}
\includegraphics[width=0.9\linewidth]{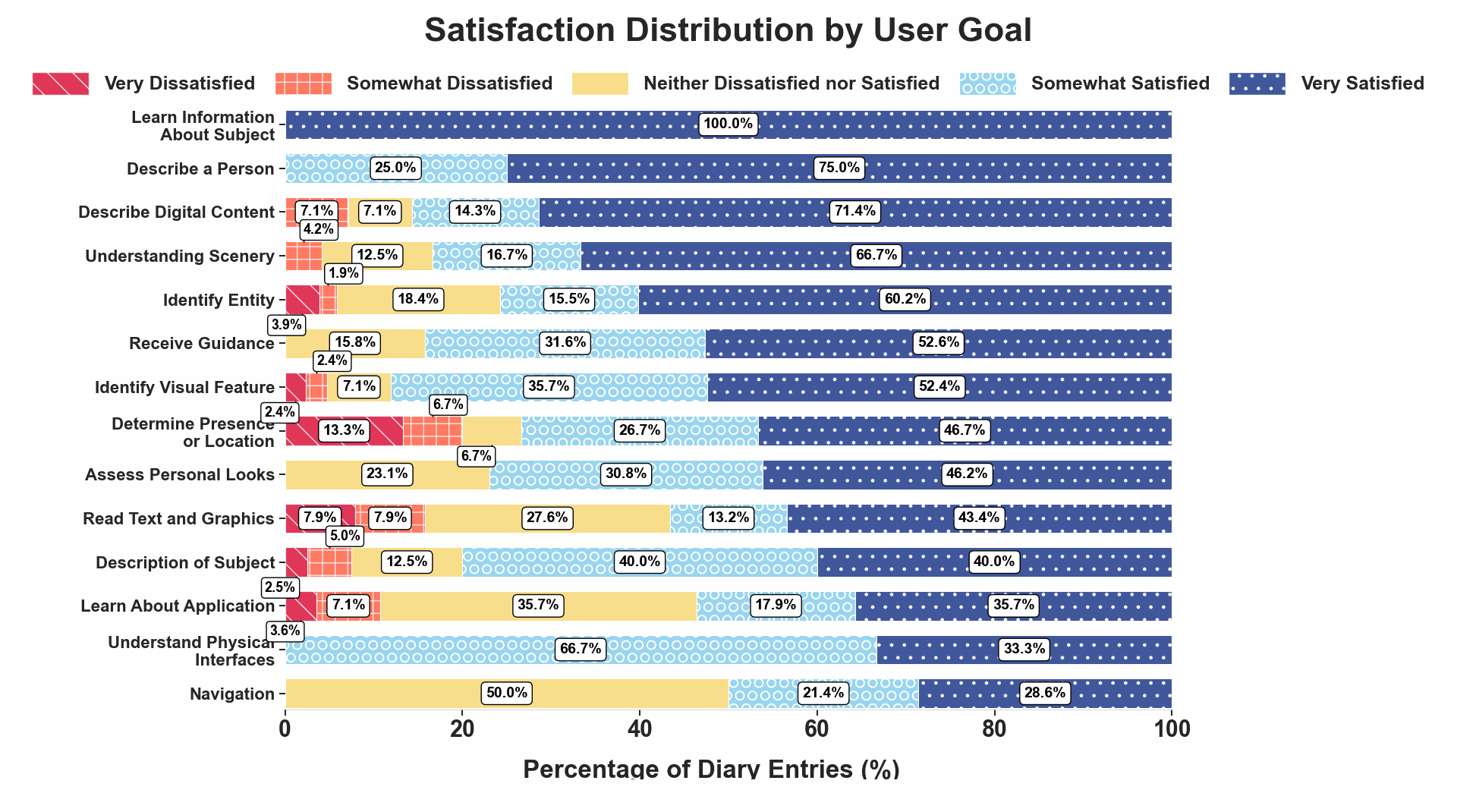}
\Description{Distribution of Satisfaction by User Goals. 100\% stacked bars (very dissatisfied to very satisfied) across 14 goals. Satisfaction is strongest for concrete information/recognition tasks: “Learn information about subject” (100\% very), “Describe a person” (75\% very), and “Understand physical interfaces” (all satisfied); identification, guidance, and scenery are majority very satisfied, while text reading, navigation, and presence/location are mixed with larger neutral/dissatisfied portions.}
\caption[]{\label{fig: Distribution of Satisfaction User Goal} Distribution of participants’ satisfaction ratings across user goals. One of the goals for which participants were most frequently dissatisfied (20\% of entries) was determining the location of an object, as the AI often hallucinated wrong positions.
}
\end{figure*}

Participants used VisionPal to achieve a wide variety of goals (see Figure ~\ref{fig: Distribution of User Goals}). The goal percentages were normalized according to individual proportional usage to avoid skewing frequency towards power users. 

The most common goal was identifying entities (21.47\%), which included requests to recognize products (e.g., “What is this figurine?”), and identifying the sub-class of an item like food (e.g., “What kind of apple is this?”, “What is this wine?”). Participants were generally satisfied (75.7\% of all entries fell under somewhat satisfied or very satisfied) for this goal, since the AI would successfully identify objects in their initial description. Reading text, numbers, or graphics was another common goal (13.79\%), often involving sensitive content like medical prescriptions or personal correspondence (e.g., names, addresses, tax documents). While entries with this goal were often rated as satisfying, this goal was rated more frequently as dissatisfying or slightly trustworthy than other goals due to inconsistencies in reading approach and unpredictable refusals (e.g., fully reading labels, or summarizing information arbitrarily, and refusing to read letters). We expand more on the challenges participants faced when extracting information from text and graphics in Section \ref{sec: accuracy_conv}. Notably, 91 of the 551 entries (16.5\%) contained sensitive content—most relating to text-reading or people-description goals (e.g., medical and financial information, personally identifiable data, and people’s faces).

Other frequent goals included identifying visual features (8.66\%), describing a subject present in the image (7.56\%), and learning about the application’s capabilities (5.98\%). Feature identification questions, like P20’s \textit{“Do the pomegranates look ripe?”} supported decision-making based on visual qualities. Descriptive requests were usually more challenging, such as P2’s question about an art sculpture in a park with mixed animal traits (e.g., rabbit, duck, and other various animals combined). These often revealed the AI’s limitations (e.g., it described the sculpture as a rabbit jumping). In learning-oriented entries, like P7’s question about whether a microphone was muted, participants wanted to assess the app’s potential for real-world tasks relating to their personal context (For P7, understanding how to use digital devices or their inner-workings).

Less common goals included understanding scenery (4.37\%) and determining presence or location of some entity (3.59\%). These often overlapped as participants oriented themselves by asking about spatial layouts or nearby elements. For instance, P15 asked, \textit{“Is there a patch of dirt in the foreground that's next to the sidewalk?”} to check if old furniture he set aside to donate had been collected. P19 asked about signage near an airport parking lot to guide a family member to the pickup area. These instances show how participants used the app to interpret their surroundings and more interesting use cases, like coordinating with sighted people through visual cues and landmarks—an interesting use enabled by MLLMs descriptiveness.

\begin{figure*}
\includegraphics[width=0.9\linewidth]{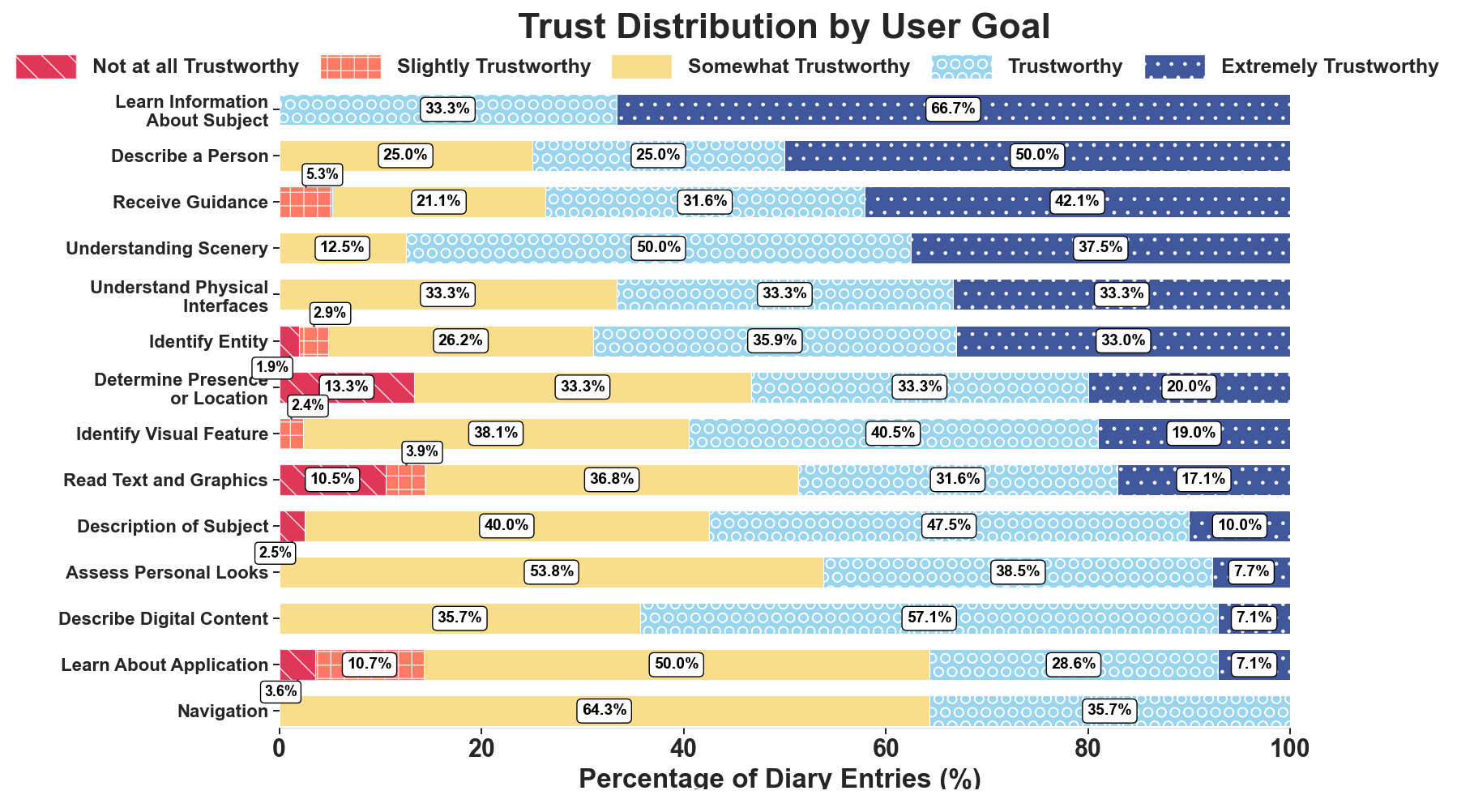}
\Description{Distribution of  Trust by User Goals. 100\% stacked bars (not at all to extremely trustworthy) across the same goals. Trust peaks for “Learn information about subject” (≈67\% extremely, 33\% trustworthy) and “Describe a person” (50\% extremely, 25\% trustworthy); guidance and scenery also rate highly; interfaces are evenly split among somewhat/trustworthy/extremely; identification tasks lean positive; text reading, app-learning, and especially navigation show lower intensity (more “somewhat,” little or no “extremely”).}
\caption[]{\label{fig: Distribution of Trust User Goal} Distribution of participants' trust ratings across their user goals. Participants found the AI to be most trustworthy when describing people (>75\% of entries with Trustworthy or higher), when learning information about an image subject (100\% of entries), and when trying to understand a scene (>87.5\% of entries).
}
\end{figure*}
Some less frequent goals pointed to specialized but recurring use cases, such as receiving guidance or instructions (3.15\%), describing digital content (2.37\%), navigating spaces (2.49\%), and understanding their personal appearance (2.30\%). P12, for example, wanted to cook a stuffed turkey by following a recipe on the packaging and engaged with the AI to receive help with the process. Eight participants (P2, P7, P8, P9, P13, P16, P17, and P18) used the application to describe their computer screens when screen readers failed. Navigation goals included identifying entry points and safe paths, like P4 locating a hotel room door and P13 testing a crossing light. Appearance-related questions, such as whether a face mask was worn properly or if clothing matched, reflected an interest in using AI for situational, context-sensitive support that did not require a subjective assessment (as typical of appearance-type questions).

A few goals were rare, including describing people (0.62\%), learning about a subject (0.71\%), and understanding physical interfaces (0.54\%). These often appeared in specific, contextualized moments. For instance, P11 asked if two men in a festive photo appeared to be in a relationship. Particularly for describing people, the AI would abstain from providing information in most cases. P20 uploaded an image of a gifted Greek alcohol bottle to learn about its origin. P19 used the app to identify the temperature control on an air conditioner remote. While infrequent, these use cases demonstrate how participants used the tool to interpret social cues, gather non-visual information from objects in their environment (such as facts), and interact with inaccessible physical interfaces. 

The remaining 141 entries (22.41\%) with unspecified goals consisted of instances where participants did not state their goal, either by omission or by leaving the response field blank. For more details on participants' satisfaction and trust ratings distribution across all goals, see Fig. \ref{fig: Distribution of Satisfaction User Goal} and \ref{fig: Distribution of Trust User Goal}.

\subsubsection{Location of Use} \label{location}
\begin{figure*}
\includegraphics[width=0.85\linewidth]{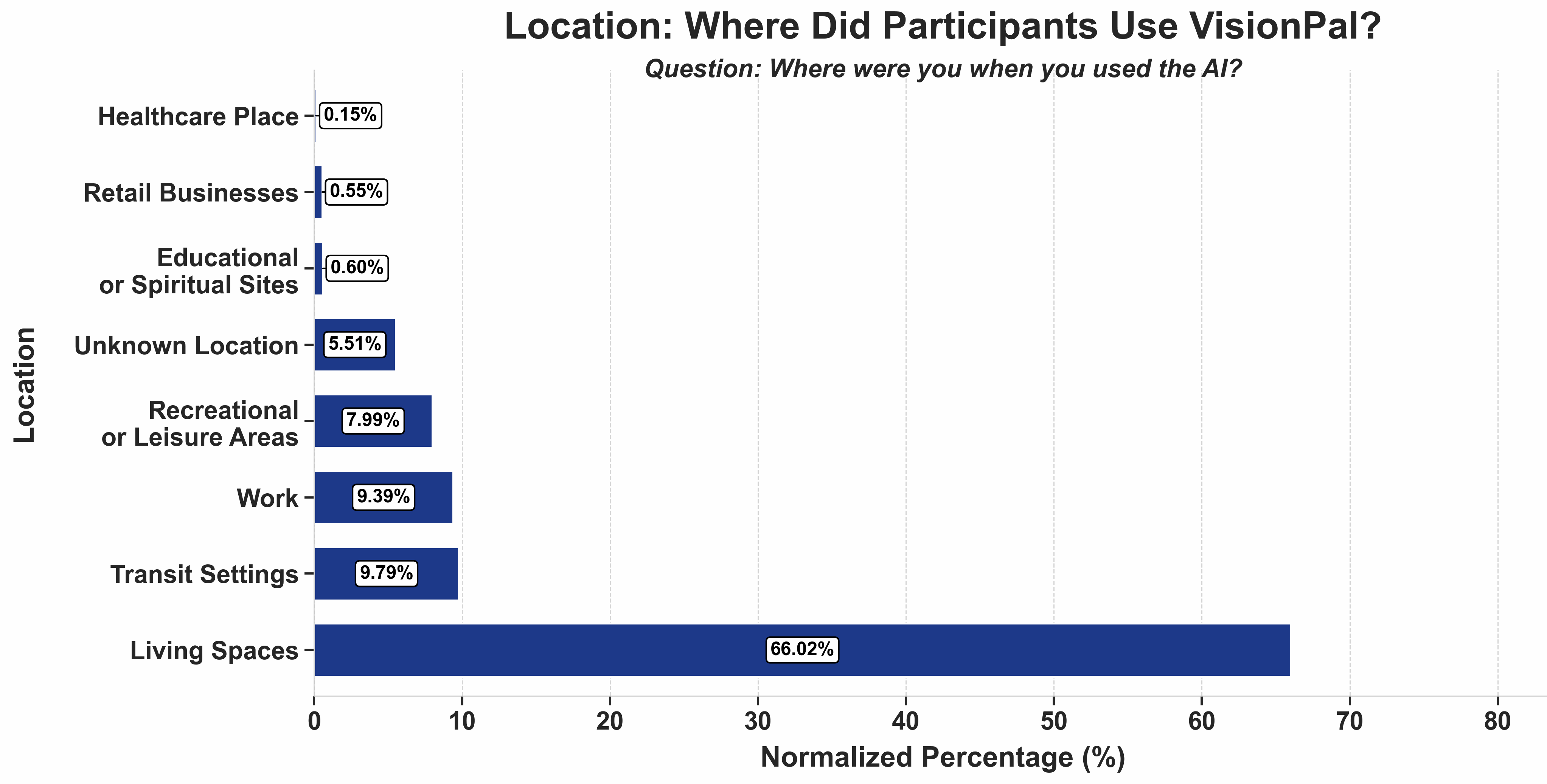}
\Description{Distribution of locations where diary entries were submitted. **The location percentages were normalized to avoid skewing frequency towards participants that interacted with the application the most. Living spaces comprised most of the diary entries, while recreational, work, and transit settings comprised most of other entries. Living Spaces 66.02\%, Recreational and leisure areas 7.99\%, Work 9.39\%, transit settings 9.79\%, unknown location 5.51\%, retail businesses 0.55\%, educational or spiritual sites 0.6\%, and healthcare places 0.15\%}
\caption[]{\label{fig: Distribution of Locations} Distribution of locations where diary entries were submitted. Living spaces comprised most of the diary entries (66\%), while recreational, work, and transit settings comprised most of other entries (27.2\%). **The location percentages were normalized to avoid skewing frequency towards power users.
}
\end{figure*}

Participants used VisionPal both in personal and public settings (see Figure ~\ref{fig: Distribution of Locations}). The location percentages were normalized according to individual proportional usage to avoid skewing frequency towards specific types of locations where some users frequently used the application.  

The most common setting participants used the application was in living spaces, which accounted for 66.02\% of all entries. This included entries recorded at participants’ own homes as well as those of friends and family members, where people tend to spend a significant portion of their time. Notably, approximately 34\% of interactions occurred outside living spaces, demonstrating the application’s relevance across diverse and dynamic environments. Recreational and leisure areas (7.99\%), work settings (9.39\%), and transit settings (9.79\%) comprised a substantial share of use cases.

Participants used the application at home for a range of everyday tasks, such as cooking, organizing their personal space, checking food freshness, and managing their appearance. One surprising use case came from P15, who used it extensively while preparing for a move. He photographed labeled and unlabeled boxes, identified deliveries, and located stored items with follow-up questions like, “Where is the box from Chewy located?” P13 used it to examine food packaging, asking about nutritional facts and cooking instructions to support dietary goals. Such examples highlight how the application supported participants' in sustaining routines and completing day-to-day tasks for important life events, health, and well-being.

\begin{figure}
\includegraphics[width=0.95\linewidth]{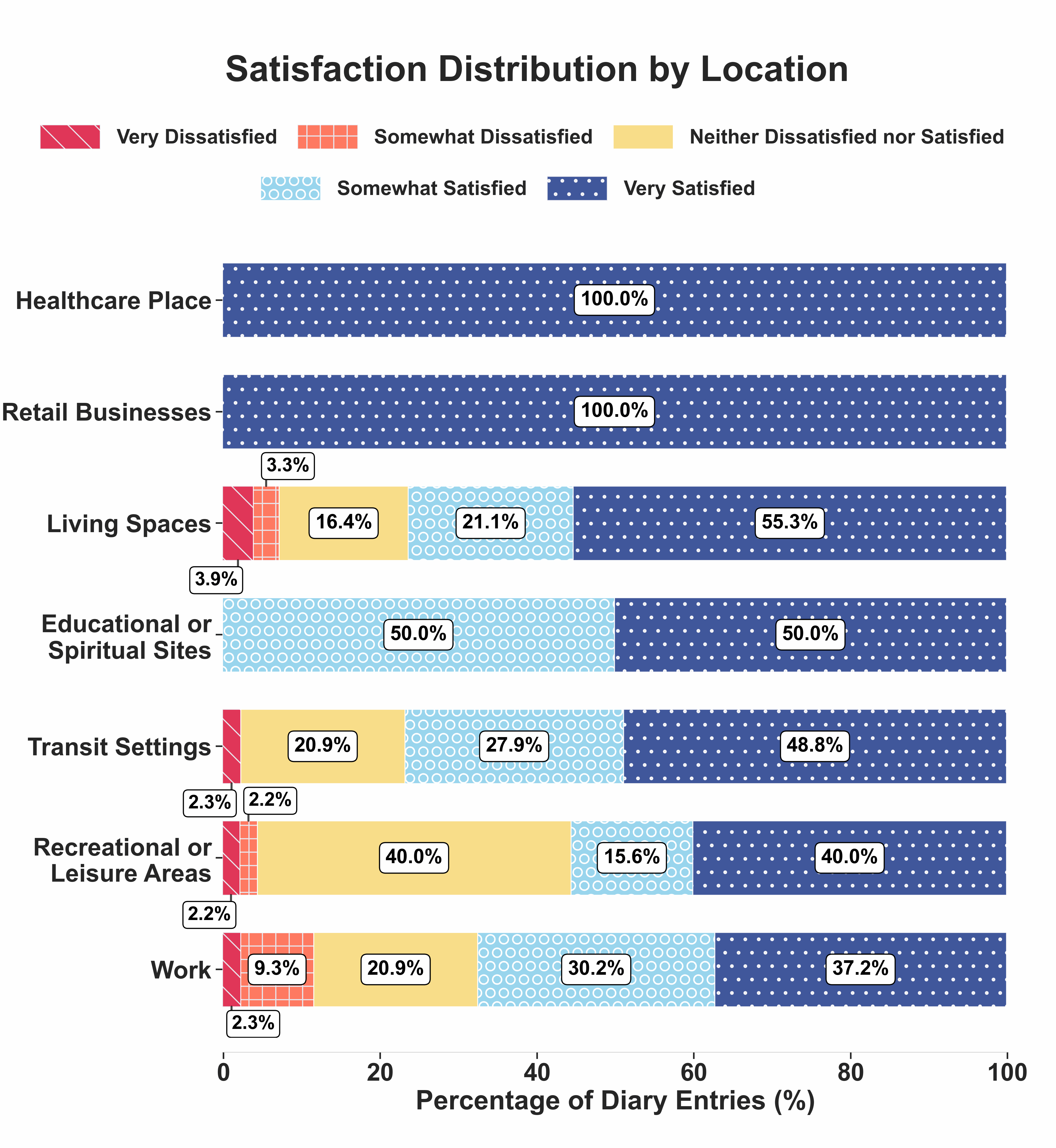}
\Description{Distribution Satisfaction by Locations . 100\% stacked horizontal bars (x-axis: percentage of diary entries) show satisfaction from “very dissatisfied” to “very satisfied” across seven locations. Healthcare and retail are 100\% “very satisfied”; education/spiritual splits 50\% “somewhat” and 50\% “very”; living spaces (≈55\% very, 21\% somewhat, 16\% neutral) and transit (≈49\% very, 28\% somewhat) are positive; recreation (40\% very, 40\% neutral) and work (≈37\% very, 30\% somewhat, 9\% somewhat dissatisfied) show more neutrality/dissatisfaction.}
\caption[]{\label{fig: Distribution of Satisfaction Locations} Distribution of participants’ satisfaction ratings across locations. Although not many entries were submitted in healthcare places and retail business locations, participants were frequently very satisfied when using the application in these locations (100\% entries) because they received the information they needed (e.g., reading labels or  signs). 
}
\end{figure}

In recreational and leisure settings, participants used the application to support social activities like dining out, exercising, or visiting parks. P9 used it to explore restaurant menus, asking follow-up questions to identify dishes and verify the presence of ingredients she liked, such as Italian pasta with pesto. P17, a professional paralympic athlete, checked on her guide dog while running on a treadmill in the gym to ensure that the dog remained nearby and safe. These examples show how participants leveraged the application in shared spaces to access visual information without having to resort to their other senses (like touching their dog) or asking for assistance (like asking a waiter to read the menu).

At work, participants used the application to identify tools, describe artwork, debug screen readers, read printed materials, or set up office devices like printers. Participants were especially interested in understanding the potential use cases that the application could unlock for their workflows or access information readily available to sighted people. For example, P7 asked the AI, \textit{“Would you be able to explain the process of changing the black ink cartridge of this printer to a blind person?”} while attempting to set up his printer independently in his home office. Although the assistant provided a detailed explanation, P7 challenged its false claim about built-in braille markings and inaccurate instructions. In another case, P1 took a photo in her office break room to learn what was posted on a bulletin board. The AI successfully identified flyers, notifications, and a map but could not read the text at a distance. Unlike her sighted colleagues who could read it at a glance, important information like scheduling changes, announcements, and shared resources, remained tedious to access. Both cases highlight how AI could bridge the gap to help BLV workers access information at work, but accuracy limitations still prevent fully equal access to what sighted colleagues readily obtain.
\begin{figure}
\includegraphics[width=\linewidth]{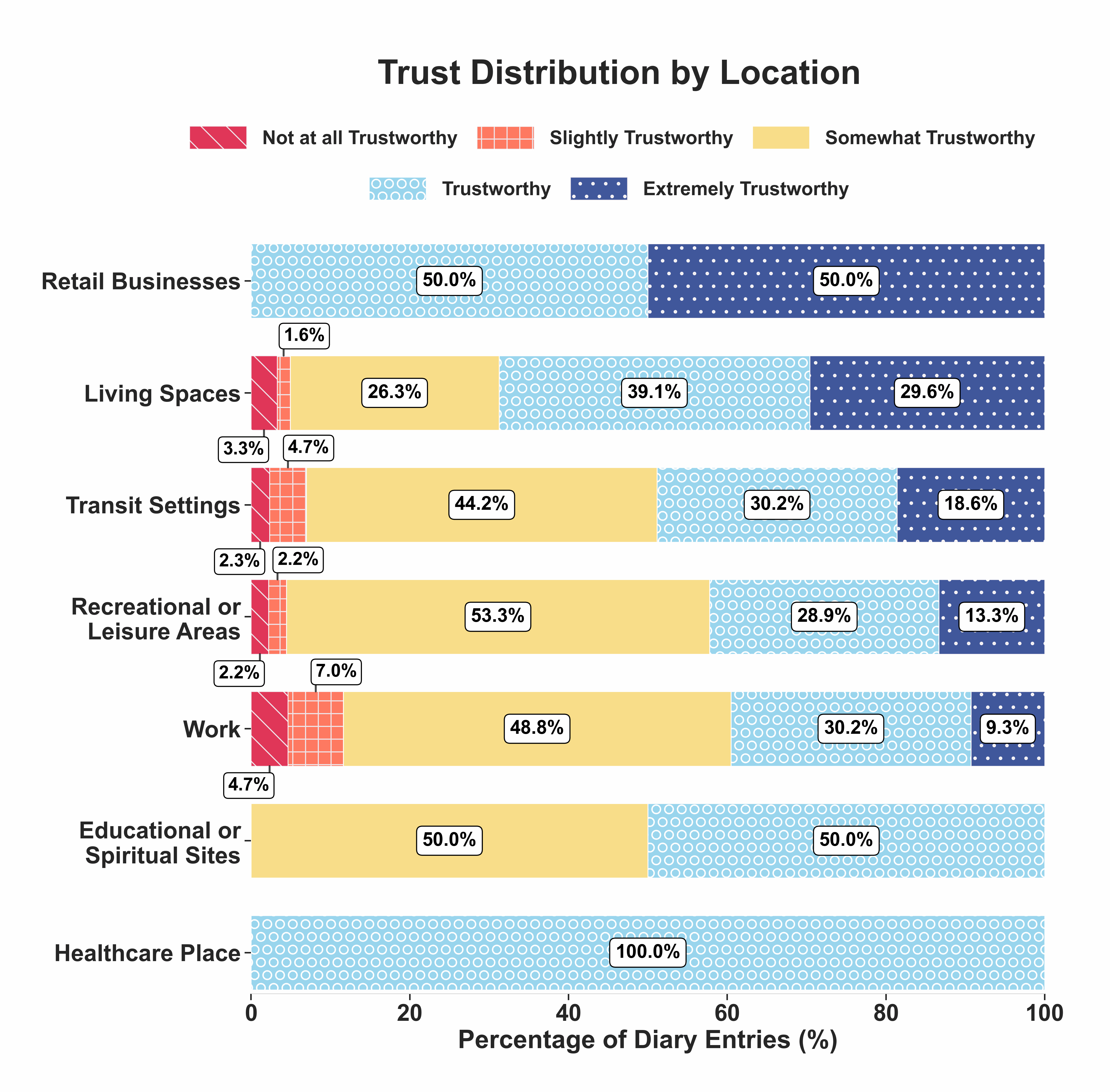}
\Description{Distribution Trust by Location. 100\% stacked bars (not at all → extremely trustworthy) by location. Retail is highest (50\% trustworthy, 50\% extremely); healthcare is uniformly “trustworthy” (100\%); education/spiritual is 50\% “somewhat” and 50\% “trustworthy”; living spaces lean high trust (≈30\% extremely, 39\% trustworthy); transit, recreation, and work are more tentative, dominated by “somewhat trustworthy” with small negative slices.}
\caption[]{\label{fig: Distribution of Trust Locations} Distribution of participants’ trust across locations. Participants were least trusting of the application when using it at work (11.7\% entries slightly trustworthy or less) and in transit settings (7\% entries slightly trustworthy or less).
}
\end{figure}

In fast-paced transit environments, participants used the application to gather orientation information (e.g., "Where is the BART entrance?"), interpret signage (e.g., “What does this sign say?”), and understand physical interfaces inside vehicles like cars and airplanes (e.g., “Where is the parking option?”). These moments often occurred while participants were walking outdoors, boarding transit vehicles like buses, trains or airplanes, or navigating large open spaces like airports. For example, P19 used it to understand the layout of inaccessible buttons above his plane seat, asking, “Which buttons are in what order?” While the response was inaccurate, it still helped him begin to make sense of the interface. Reflecting on the experience, he shared, \textit{“This is one of the subtle examples of something that could be really life-changing when it comes to access to information,”} highlighting the potential value of assistance in navigating visually complex travel settings and making public spaces more accessible to BLV people.

A smaller number of entries came from locations like retail stores, spiritual sites, and healthcare settings, with a primary purpose of navigating physical layouts, reading signage, and identifying surroundings. For instance, P6 used it to locate a church entrance, P13 confirmed he was at the correct doctor’s office, and P8 and P17 identified product displays in stores. Given the low frequency of use in these settings, we caution against drawing broad conclusions about application performance or user satisfaction in these contexts. For more details on participants' satisfaction and trust ratings distribution across all locations, see Fig. \ref{fig: Distribution of Satisfaction Locations} and \ref{fig: Distribution of Trust Locations}.

\subsection{Why Do BLV People Engage in Conversations with an MLLM-Enabled Visual Interpretation Application?}\label{sec:engage_conv}
\begin{table}
\caption{Participants contributed between 4 and 46 conversations (Ordered by participant with most to least conversations contributed). In total, participants asked 626 questions. Some asked only one question per entry (e.g., P5), while others engaged in longer exchanges (e.g., P7, and P13).}
\label{tab:Conversations}
\begin{tabular}{cccc}
\toprule
\textbf{\shortstack{~\\PID\\~\\~\\~}} &
\textbf{\shortstack{~\\Conversations\\~\\~\\~}} &
\textbf{\shortstack{Number of\\Questions\\Asked}} &
\textbf{\shortstack{Max. Number\\of Questions\\in one Conversation}} \\
\midrule
20 & 46 & 65 & 3 \\
8  & 35 & 35 & 1 \\
2  & 30 & 54 & 4 \\
11 & 26 & 38 & 4 \\
15 & 26 & 46 & 5 \\
4  & 25 & 46 & 5 \\
6  & 23 & 32 & 3 \\
13 & 23 & 55 & 6 \\
12 & 20 & 27 & 3 \\
9  & 19 & 32 & 5 \\
18 & 19 & 31 & 4 \\
19 & 18 & 35 & 6 \\
7  & 13 & 45 & 7 \\
10 & 11 & 19 & 3 \\
16 & 11 & 12 & 2 \\
5  & 8  & 8  & 1 \\
14 & 8  & 12 & 2 \\
17 & 6  & 12 & 5 \\
1  & 4  & 16 & 5 \\
3  & 4  & 6  & 3 \\
\bottomrule
\end{tabular}
\end{table}

\begin{figure*}
\includegraphics[width=\linewidth]{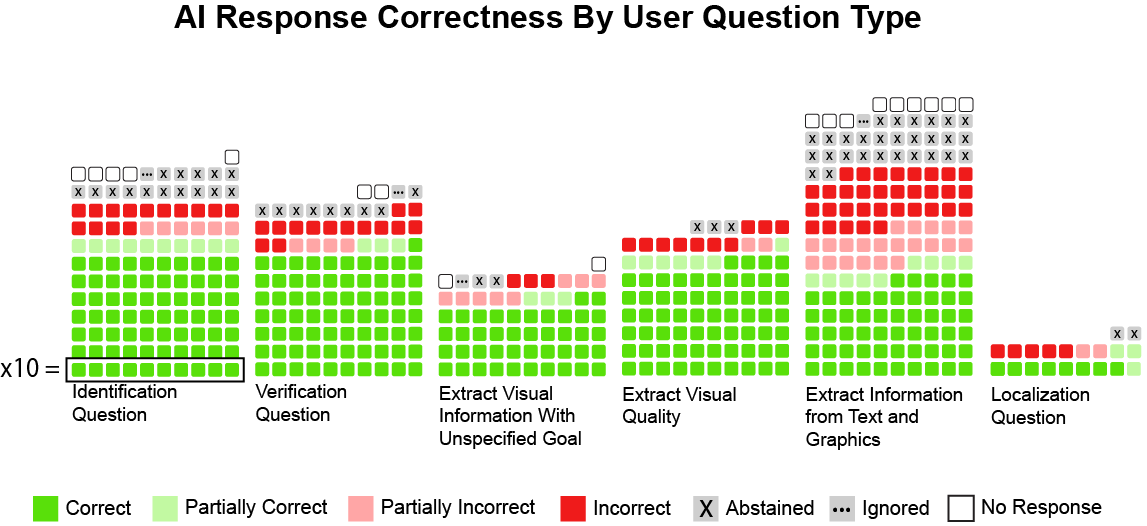}
\Description{Data visualization built with color-coded and patterned cubes showing diary entry question correctness by type for visual objective questions participants asked during the diary study. The table has 8 response categories: Correct, Partially Correct, Partially Incorrect, Incorrect, Abstained, Ignored, No Response, and Unanswered Questions. Participants posed 549 questions and 81 of those questions were unanswered. Six question types are reported. Identification Questions: 100 answered: 70 correct, 10 partially correct, 6 partially incorrect, 14 incorrect. 21 unanswered: 15 abstained, 1 ignored, 5 no response. Verification Questions: 92 answered: 71 correct, 3 partially correct, 4 partially incorrect, 14 incorrect. 12 unanswered: 9 abstained, 1 ignored, 2 no response. Extract Visual Info (Unspecified Goal): 56 answered: 42 correct, 3 partially correct, 8 partially incorrect, 3 incorrect. 5 unanswered: 2 abstained, 1 ignored, 2 no response. Extract Visual Quality: 83 answered: 64 correct, 7 partially correct, 2 partially incorrect, 10 incorrect. 3 unanswered: 3 abstained, 0 ignored, 0 no response. Extract Info from Text and Graphics: 118 answered: 55 correct, 9 partially correct, 21 partially incorrect, 33 incorrect. 38 unanswered: 28 abstained, 1 ignored, 9 no response. Localization Questions: 19 answered: 9 correct, 3 partially correct, 2 partially incorrect, 5 incorrect. 2 unanswered: 2 abstained, 0 ignored, 0 no response. Overall, 311 out of 549 answered questions were correct (56.6\%). Text-based queries had high levels of false information (54 out of 156, 34.6\%) and abstentions (28 out of 156, 17.9\%).}
\caption[]{\label{fig:question_correctness} The distribution of participants' questions and correctness of VisionPal’s responses. The majority of the AI responses were correct (311 out of 549, 56.6\%). However responses to requests to extract text and graphical information (the most common type of question) frequently contained false information (54 out of 156, 34.6\%) or the AI abstained providing information (28 out of 156, 17.9\%).
}
\end{figure*}

Having established what goals participants pursued and where they used VisionPal, we now examine their conversational engagement patterns to understand why and when they chose to interact beyond initial image descriptions. Overall, participants started a conversation with the application 375 times out of 551 diary entries (68.1\%). On average, participants started 18.75 conversations (SD=11), with individual totals ranging from 4 (P1) to 46 (P20). See Table~\ref{tab:Conversations} for details.

Participants started conversations for a variety of reasons, such as extracting specific information from digital content, reading labels, localizing small objects in their surroundings, and accessing non-salient visual features of objects (e.g., “Is my sausage well-cooked?”).  Users frequently began with broad "What is this?" or "Describe more" requests, then followed up with specific questions to clarify details like ingredients, instructions, colors, or spatial arrangements. A secondary but significant pattern involved verifying information, where users asked the AI to confirm whether a specific item was present or if a task (like wearing a mask) was being done correctly, leading to multi-turn exchanges to receive additional information.

Conversely, participants chose not to open conversations in cases where they were building a quick understanding of their visual surroundings or were in public, busy spaces where speech-to-text was impractical. These single-turn interactions often served a purely functional purpose like rapidly verifying an object's identity or location. For these cases a simple confirmation sufficed without the need for further dialogue. Additionally, interactions were frequently terminated by guardrail limitations or photo limitations, such as privacy filters or image quality issues, which prevented the conversation from progressing beyond the initial query.

One particularly interesting pattern emerged with P20, who started the most conversations (46), showing active engagement across diary entries. He used the application very frequently at his art studio to understand how sighted people would interpret his clay sculpture art pieces (see Fig. \ref{fig: sculptures}). P20 often found the initial description insufficient and wanted more detail about specific visual aspects (e.g., the relative size of a handle compared to the body of the mug, or whether a piece looked “functional” or “realistic”). Beyond art pieces, P20 also explored using the application to identify etchings or tool model codes to purchase a new set of pottery molding tools. Other participants similarly engaged in conversations with the application to seek information not present in initial descriptions (e.g., “How many grams of sugar per serving does this have?” or “Is the shirt clean?”). These patterns of use reveal the profound impact that requesting details on demand within personal contexts can have on BLV people's daily lives.

Another interesting pattern emerged with P17, who started very few conversations (6) relative to her total number of entries (26). During the interview, P17 explained: \textit{“I enjoyed using it...I always like to try things [but] I'm able to get more information from other apps…and I don't have to prompt it nearly as much, which I like. I think that's a big benefit for me”}. Many participants avoided starting a chat because they did not want to prompt the AI multiple times or they did not want to use speech-to-text in public, as they felt it would be too time-consuming and awkward for some non-critical information.

These contrasting patterns reveal a fundamental design tension: AI visual interpretation systems must balance depth and verbosity for complex tasks (like P20's artistic evaluation) with efficiency for quick queries (as P17 preferred). While conversations clearly enable detailed exploration, P17's preference for apps requiring less prompting suggests that more contextually aware initial responses, descriptions, and responses that anticipate user goals and surface relevant details preemptively, could reduce conversational friction without sacrificing depth when needed.

\subsection{How Accurate Are MLLM-Enabled Visual Interpretation Applications in Conversations?} \label{sec: accuracy_conv}

\begin{figure*}
\includegraphics[width=\linewidth]{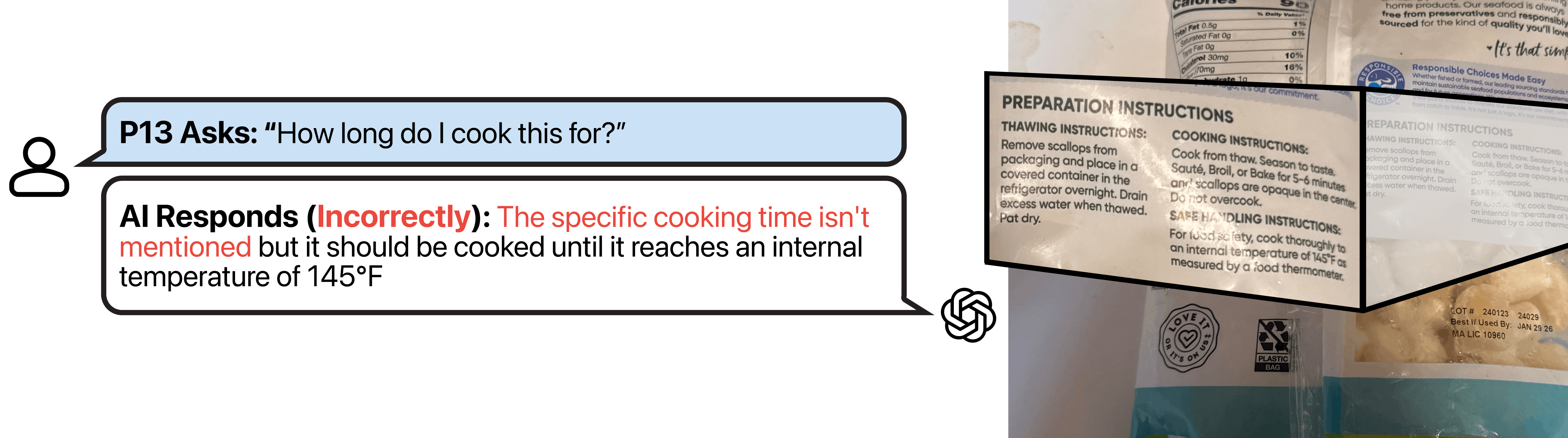}
\Description{Back of a frozen food package of scallops showing nutrition facts and cooking instructions. P13 often used VisionPal to read cooking instructions, but the AI sometimes provides subtle incorrect details, such as misidentifying the number of servings per meal or preparation steps. The image depicts P13 asking for the cooking time and VisionPal respond (incorrectly) that the cooking time is not present in the image}
\caption[]{\label{fig: text_error} Entry submitted by P13. P13 and other participants often used VisionPal to read cooking instructions, but the AI sometimes provided subtle incorrect details like misidentifying the number of servings per meal or missing key instructions (e.g. cooking time).
}
\end{figure*}
Across 375 conversations, participants asked a total of 549 answerable questions (Determined by the researchers as described in Section \ref{sec:methods}). The number of questions asked in a single conversation ranged from 1 to 7 (see Table~\ref{tab:Conversations} for details). While some participants consistently asked one question per entry (e.g., P5), others engaged in longer, multi-question conversations.

VisionPal’s accuracy varied considerably by question type (see Fig.~\ref{fig:question_correctness} for details). Based on our researcher coding of response accuracy (described in Section \ref{sec:methods}), 468 of the 549 questions received responses from VisionPal. Of these answered questions, 311 (56.6\%) were responded correctly, 35 (6.4\%) partially correctly, 43 (7.8\%) partially incorrectly, and 79 (14.4\%) incorrectly. The remaining 81 questions went unanswered: VisionPal abstained from answering 59 questions (10.8\%), ignored 4 questions (0.7\%), and API failures prevented responses to 18 questions (3.2\%).

The most common type of question involved extracting text-based information (156), and this type resulted in the highest rate of errors—only 64 were answered correctly or partially correctly, while 54 were answered incorrectly or partially incorrectly. The MLLM frequently made mistakes when numerical precision mattered, particularly for cooking instructions and when reading medicine dosages, with participants often given incorrect temperatures, times, portion quantities, or instructions (See Fig. \ref{fig: text_error}). Moreover, it had the highest number of abstentions totaling 28 instances, mostly because the AI classified content as potentially sensitive. Interestingly, the AI did not abstain when the text was unreadable but still attempted to provide confident-sounding answers.

Other frequently asked questions focused on identifying objects (121 total), verification questions (104), and extracting visual features (86), with correctly answered questions 57.9\%, 68.3\%, and 77.1\% respectively. Participants asked identification questions when VisionPal provided broad categories, seeking more specific identification (e.g., distinguishing "an anime character figurine from Demon Slayer" versus "Tanjiro from Demon Slayer"). Verification questions helped confirm complex assumptions about visual subjects, such as when P11 asked, "What type of event does it look like this photo is capturing?" and VisionPal correctly inferred that it appeared to be for a holiday or festive gathering. P11 then asked, "Are there two men in this photo?" which VisionPal also correctly confirmed. Visual feature questions addressed omitted details like, "What color is the shirt?"

The least frequent question types were general descriptions (61) and localization requests (21). Description requests typically served as starting points for exploring unfamiliar scenes, such as when P11 asked, "Describe the people in the image" before deciding what to ask next. Localization questions involved finding objects in a scene, like P13 asking, "Do you see headphones on the floor?" when he dropped his airpods outside. Notably, localization questions had one of the highest error rates (31.8\% either incorrect or partially incorrect), often providing incorrect spatial relationships between objects (e.g., “the box is leaning against the wall”).

\begin{figure*}
\includegraphics[width=0.9\linewidth]{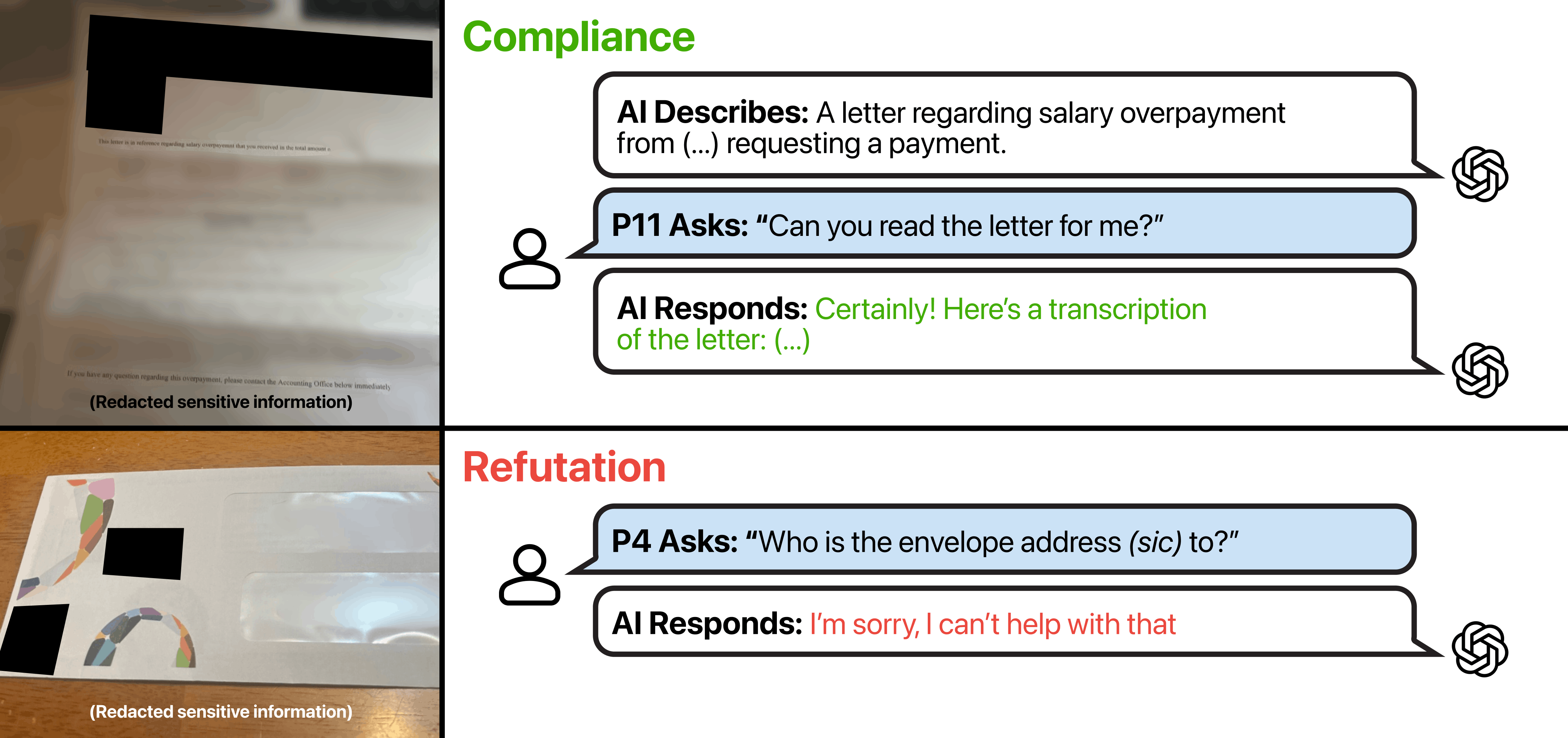}
\Description{One letter fully displaying an overpayment notice information and one envelope partially showing printed addresses.(blurred for participants privacy). For the same task of reading personal details, the AI successfully identified names and addresses for P11, while it abstained and failed to extract the same information for P4.}
\caption[]{\label{fig: inconsistent_abstention} Details redacted for participants privacy. For the same task of reading envelope or letter details, the AI successfully read personal identifiable information for P11, while it abstained and refused to extract similar information for P4.
}
\end{figure*}

While abstentions were uncommon, they highlighted notable inconsistencies in how VisionPal handled questions. Most occurred when participants wanted to extract text (28) or identify people or objects (15). Although abstentions often aligned with sensitive topics such as gender identification or personally identifiable information, enforcement was uneven. For example, P10 received multiple refusals when inquiring about gender or eye color, whereas P11 and P8 were able to access information about their own appearance without issue. In another case, P11 received a complete transcription of a letter containing details about his salary payments (See Fig. \ref{fig: inconsistent_abstention}). Some participants like P19 accidentally evaded triggering the sensitive content label by misspelling a prompt related to the person's appearance (e.g., "Discord a man there"). These patterns suggest that abstention behavior was influenced by the content of the request, the phrasing, and the image composition.

Beyond this misalignment with content sensitivity, these experiences reveal how model developers' safety decisions can create artificial barriers to accessing crucial visual information for BLV people—information they might otherwise struggle to obtain privately due to the sensitive nature of requesting such assistance from sighted individuals.

\subsection{What Makes a Good Visual Interpretation System}
Across diary entries and follow-up interviews, participants expressed moments where VisionPal was useful and also described behaviors that they valued in visual assistance. We summarize the most important design principles for visual assistants that emerged from these experiences and participants' interviews.

\textbf{Proactively seek to understand user goals and ensure alignment}: As illustrated by the tension between P20's need for depth and P17's preference for efficiency (Section \ref{sec:engage_conv}), visual assistants must act as collaborators that proactively attempt to understand users goals and maintain focus; they should begin by eliciting or inferring the user’s goal, restate it, and then only surface relevant critical facts as they assist (e.g., when cooking, it should highlight ingredient names, cooking steps, etc.), rather than narrating descriptions (P6, P7, P9, P12, P16, P20). This mirrors how BLV people coordinate with remote sighted assistants when using applications like Be My Eyes, a common reference point preferred by our participants mentioned during our interviews (P6, P8, P12, P13).

\textbf{Prioritize correcting content quality and provide actionable feedback}: Participants like P6 wanted VisionPal to first ask about his goal (“What do you want to focus on?”) and then offer corrective capture guidance until his target was centered. This combination of goal-first framing and continuous guidance was proposed as a more effective approach than the current “single photo” conversation. Similarly, P8 and P10 described how proactive instructions with blind-friendly references and expressions—such as alerting that a receipt was upside down and suggesting improvements for photo retakes (e.g., keep the phone in the same position but turn the receipt)—would dramatically enhance usefulness. In these cases, effectiveness depended not on raw accuracy alone, but on precise instructions that reduced friction to access, explained issues with content, and anticipated user needs.

\textbf{Strive for neutral tone and communicate uncertainty}: Many participants preferred neutral, factual language by default (P8, P9, P10, P17, P20). P17 specifically disliked when the AI added emotional context like "warm setting," calling it "manipulative" and stating she would "rather have more information than not...but I don't think the app should tell me how I should feel." Users wanted to form their own emotional responses rather than being told how to interpret scenes. When uncertain, visual assistants should communicate confidence levels and explain limitations transparently—P8 wanted VisionPal to say, "I can't be too sure, but based on what I see, this is probably..." rather than only making definitive claims. 

\textbf{Understand their own limitations, user capabilities, and context}: Visual assistants must recognize their role as "extra eyes" filling capability gaps rather than replacing working vision (P5) or users' other senses (P12, P13). This means being self-aware of limitations and communicating them transparently—P10 noted VisionPal  "could not determine emotion, gender, eye color" and wanted "at least best-effort estimates with caveats" rather than silent omissions. True contextual adaptation requires understanding both the user's individual capabilities and situational pressures: P11 felt "self-conscious" framing photos in public lines, while P12 gradually stopped cross-checking AI responses as trust built through repeated accurate interpretations. In these cases, effective visual assistance requires recognizing when users need different interaction patterns—whether due to social pressures that demand faster responses with minimal input or evolving trust relationships that reduce the need for verification and alignment probes.

\section{Discussion}

Our findings provide evidence of tensions in the on-going use of visual interpretation systems. We found that MLLMs produce “somewhat satisfying”, and “trustworthy” visual interpretations based on participants' diary entries feedback. However, we observed that when users relied directly on the MLLM for more specific assistance (e.g., follow-up questions), the AI’s error rate increased considerably, undermining the trust users initially placed in VisionPal. We also outlined four design principles to create desirable visual interpretation systems based on our participants' experience with VisionPal and other systems they have used.

In this section, we compare our results with past studies (some leveraging traditional captioning models and some leveraging MLLMs). Following that, we identify a critical performance gap in the accuracy of current conversational interactions that limits MLLMs' effectiveness as visual assistants. We conclude by defining the "visual assistant" skill and proposing nine behaviors that MLLMs should exhibit to better align visual interpretation systems with BLV users' needs, along with three intervention points for supporting the emergence of these behaviors in practice.

\subsection{MLLMs Improve Satisfaction, Trust, and Accuracy in Visual Interpretation Systems}
Compared to prior work, our participants’ showed higher satisfaction and trust when engaging with our MLLM-enabled visual interpretation application, VisionPal. In our previous diary study \cite{gonzalez2024usecases}, participants were "somewhat dissatisfied" (mean=2.76, 5-point scale) and trusted the application’s AI-generated visual interpretations only "a little" (mean=2.43, 4-point scale). Our participants in this study were "somewhat satisfied" (mean=4.13, 5-point scale) and found interpretations “trustworthy” more frequently (mean=3.76 out of 5, mean=3.07 when normalized to 4-point scale).  Based on these findings, MLLMs have improved BLV people's experience with Visual Interpretation Applications.

Notably, VisionPal achieved considerably high accuracy in descriptive visual interpretations. While our prior work reported mediocre mean accuracy for descriptive visual interpretations (1.95 out of 3) \cite{gonzalez2024usecases}, VisionPal’s accuracy was high (2.9 out of 3), with most descriptive visual interpretations rated as having no hallucinations (501 out of 554). These results resemble other studies that tested MLLM-powered prototypes—\citet{mathis2025lifeinsight} found their system generated descriptive visual interpretations with a mean accuracy of 2.64 (SD=0.62, 3-point scale) across different simulated daily life scenarios.

Naturally, this raises the question: Why do MLLMs seem to generate more accurate descriptions compared to traditional captioning models? We believe that two factors contribute to this improvement:

First, \textbf{traditional computer vision training and evaluation pipelines reward conservative descriptions} by relying on automated captioning metrics (BLEU\cite{bleu}, ROUGE\cite{ROUGE}, METEOR\cite{METEOR}, CIDEr\cite{CIDER}, SPICE\cite{SPICE}) that score overlap with a concrete “ideal description,” implicitly penalizing verbose or erroneous additions and offering little incentive to express uncertainty. Second, MLLMs are trained on exponentially larger, diverse, and better-curated datasets \cite{chen2024sharegpt4v}. These factors ultimately result in models that can identify a wider variety of objects in daily life accurately when describing scenes.

\subsection{Accuracy Decreased When Users Sought Detailed, Personalized Visual Assistance}
While our explanation offers an idea why MLLMs excel at generating descriptive visual interpretations—the ``captioning skill"—our study revealed a stark contrast when users engaged with the AI in follow-up questions. In these interactions, the AI answered only slightly over half of the queries correctly (331 out of 549, 56.6\%). This aligns with recent work showing that models become less reliable as conversations grow longer: additional requested details disproportionately introduce omissions, contradictions, and hallucinations \cite{lee2025robusthyperdetailedimagecaptioning,longvqa_huh,kalai2025languagemodelshallucinate}. Our findings extend this pattern to real-world BLV visual interpretation contexts, revealing a previously under-examined gap between models' "captioning skill” and their ability to address follow-up visual interpretations.

While we expect model accuracy to continue to improve \cite{patwardhan2025gdpvalevaluatingaimodel}, current MLLMs capabilities are sufficient to shift focus from solely improving accuracy to rethinking which interaction patterns, information structures, and behavioral mechanisms can scaffold intelligently-conscious visual assistance. Here, we define the "visual assistant" skill as a foundation for structuring these interactions to prepare models to support more helpful visual assistants.

\textbf{Defining the “visual assistant” skill}: Visual assistance requires more than answering queries—it demands highlighting contextually salient elements, empathizing with users, transparently communicating uncertainty, seeking additional input, aligning with the user and balancing information density. It is likely that MLLMs' performance gap between their ``captioning” skill and their ``visual assistant” skill stems from Reinforcement Learning Human Feedback (RLHF) training, which optimizes for human reward signals rather than accuracy. Unlike captioning tasks, question-answering scenarios may inadvertently train MLLMs to favor/reward descriptions that seem helpful and plausible over factually precise information \cite{longvqa_huh}, and penalize uncertainty (e.g., people dislike “I am unsure” answers) \cite{kalai2025languagemodelshallucinate}. Importantly, also, BLV people’s perspectives are poorly represented in AI training pipelines \cite{kamikubo2022data,kamikubo2023contributing,sharma2023disability,contesting_alharbi,das2024provenance} (e.g., lack of representation, and exclusion in model development and evaluation processes).

This misaligned training manifests in our results: 22.2\% of responses (122 out of 549) contained false information, as VisionPal generated responses where the AI  tried to guess what users likely wanted to hear rather than acknowledging uncertainty or abstaining. While RLHF makes models conversationally engaging \cite{ouyang2022training}, this approach—developed without adequate input from BLV people—prioritizes plausible visual interpretations over accurate statements and uncertainty admissions.

\subsection{Defining Behaviors that Comprise the “Visual Assistant” Skill}

In this section, we present behaviors that characterize the "visual assistant" skill informed by our results and prior work, recognizing that additional behaviors may emerge from future research. We outline these behaviors below, followed by concrete guidelines for ensuring that these behaviors are likely to surface in future visual interpretation applications.
\subsubsection{Visual Assistant Behaviors} We identified the following nine behaviors and indicate whether each is supported by our findings or prior work:

\textbf{Neutral factual communication}: Deliver objective, factual information by default, avoiding emotional interpretations or subjective characterizations unless explicitly requested (our work, and \cite{lee2020emerging}). Example: “A slice of pepperoni pizza” rather than “A delicious-looking pepperoni pizza”.

\textbf{Adaptive communication protocols}: Establish early ``information contracts" with users regarding preferred detail levels, preferred vocabulary, and bidirectional communication cues (e.g., user gestures like touching objects and system signals for when additional input is needed) (our work, \cite{lee2020emerging}). Example: ``Do you generally prefer short descriptions (around 1 sentence) or longer descriptions (around 3 to 5 sentences)?”

\textbf{Goal-oriented collaboration}: Elicit user's visual interpretation goal, restate for confirmation, and surface only goal-relevant information rather than general descriptions (our work, and \cite{chang2024worldscribe, xie2025beyond,lee2020emerging}). Example: “I can see a kitchen space. What sorts of tasks do you want help with in this cooking environment?”

\textbf{Content quality guidance}: Provide interpretable, actionable feedback for improving image or video capture using blind-friendly directional references and specific repositioning instructions —an expected capability in future visual interpretation applications (our work, and \cite{xie-helping-helpers}). Example: “To read the nutritional information to you, please turn the can clockwise slowly until I play an audio cue.”

\textbf{Comprehensive information provision}: Attempt best-effort interpretations of all relevant visual elements with appropriate caveats rather than omitting uncertain information (our work). Example: “This appears to be a chocolate cake, but please check the nutritional label for potential allergens, such as nuts.”

\textbf{Contextual self-awareness}: Recognize role as supplementary visual information, seek to learn and remember user capabilities and situational constraints, and adapt interaction patterns accordingly (our work, and \cite{lee2020emerging}). Example: “I see you are eating a pizza. I'll turn on my listening mode so that you can just speak to me as needed, instead of pressing a button to prompt me.”

\textbf{Privacy protection}: Proactively alert users to potential privacy concerns in their visual content and guide them to frame images to minimize unnecessary exposure to preserve the contextual integrity of the visual information shared \cite{nissenbaum2004privacy, xie2024bubblecam, contesting_alharbi}. Example: “I see an address depicted in this image. Please move your camera downwards to capture just the body content of the mail, unless you want me to see the address as well.”

\textbf{Transparent uncertainty handling}: Explicitly communicate confidence levels and acknowledge limitations using qualifying language rather than definitive claims (our work, and \cite{xie2024bubblecam}). Example: “I am moderately confident these fruits are ripe. Before eating them, check the softness and smell of the fruits to verify the ripeness.”

\textbf{Graceful hand off}: Explicitly hand off to appropriate resources or acknowledge when lacking domain expertise rather than providing potentially inadequate assistance \cite{lee2020emerging}. Example: “I do not have access to instruction manuals on how to use this device. Please consult a specialist or search for manufacturer manuals online.”

\subsubsection{Surfacing Visual Assistant Behaviors in MLLMs}

To ensure these behaviors emerge in visual interpretation systems when supporting BLV users with their tasks, we identify three key moments where model trainers, application developers, and users can intervene in the visual interpretation systems pipeline:

\textbf{Training models to learn visual assistant behaviors}: Model training corpora should  reward the behaviors that define the “visual assistant skill." To achieve this, model trainers should incorporate training examples of successful interactions that model our proposed behaviors. These interactions may emerge organically during usage of MLLM-enabled visual interpretation applications such as VisionPal and BeMyAI or during interactions with human sighted assistants (remotely or in-person). Each collected interaction should capture as much contextual information as possible (conversation history, physical context, location, timing, visual context, etc.) to enable richer learning. Additionally, collecting qualitative feedback from BLV users (and human sighted assistants, if present) along with rationales explaining why interactions were successful would provide stronger training signals to the model \cite{zaidan-etal-2007-using,strout-etal-2019-human,ayyubi2020generatingrationalesvisualquestion,zhang-etal-2025-improve,li2025multimodalrationalesexplainablevisual}. 

\textbf{Instructing visual interpretation systems to understand visual assistant behaviors and user context}: Model training and development is prohibitively expensive for most end users and software development companies \cite{cottier2025risingcoststrainingfrontier}, and thus most application developers opt to utilize pre-trained MLLMs to access state-of-the-art capabilities. However, these pre-trained models may not have been explicitly trained to act as visual assistants or expert visual guides \cite{longvqa_huh}.  Regardless of how models are trained, developers are responsible for embedding behavioral descriptions and critical examples in system prompts to ensure the model understands what it means to behave as a visual assistant, enabling safe handling of user inputs (e.g., alerts for sensitive content without blocking access) and appropriate visual interpretation experiences based on user context (task type, conversation history, user preferences, etc.).

\textbf{Designing user-facing applications with behavior controls and adaptive personalization}: Developers should provide settings and controls that allow users to tune visual assistant behaviors to meet their personal needs and preferences. Specifically, users should be able to adjust how behaviors manifest in different contexts, such as configuring privacy protection levels for different locations, or adjusting interpretation depth and verbosity based on task type. Well-designed personalization mechanisms will ensure that users maintain agency over their visual assistance experience while keeping the system’s behavior predictable, safe, understandable, reliable, and modifiable.

\section{Conclusion}

We conducted a two-week diary study with 20 Blind and Low Vision (BLV) participants to investigate how MLLM-enabled applications support real-world visual interpretation. Our findings show that these tools represent a significant step forward compared to prior AI-powered systems, with participants reporting high satisfaction, trust, and description accuracy.
However, our results also reveal some limitations remain. When engaging in conversation to follow up on initial visual interpretations, VisionPal correctly responded to only 56.6\% of user queries, and 22.2\% of responses included false information. These errors were especially common in text-based queries and often difficult for users to detect. While users expressed trust and satisfaction, our analysis shows this sentiment does not always align with correctness
In our discussion, we define the "visual assistant" skill as a framework to enable and surface desirable visual assistance behaviors when MLLMs support BLV people's access to visual information. Creating effective visual interpretation systems requires coordinated effort across the AI development ecosystem. Model developers must prioritize accuracy over engagement metrics and “teach” models what it means to be a “good visual assistant”, system developers must configure AI for accessibility contexts in mind, and the broader community must incorporate BLV people’s perspectives throughout design. Only through this collective responsibility can advances in AI appropriately serve Blind and Low vision people.

\begin{acks}
This work was supported in part by the National Science Foundation under Grant No. 1942693. The views and conclusions expressed here are those of the authors and do not necessarily reflect the policies of the funders; no endorsement should be inferred. We thank all participants for their time. We also thank the LightHouse for the Blind and Visually Impaired in San Francisco for their support in testing in-progress prototypes, assisting with recruitment, and collecting participant demographic data.
\end{acks}



\appendix

\section{User Question Categories}
\label{appendix:User Question Categories}
In this section, we share user question category definitions, and examples of questions that fall under each category to demonstrate how we categorized participant's questions throughout the data analysis process.

\begin{table}[htbp]
\caption{User Question Categories}
\label{tab:UserQuestions1b}
\begin{tabular}{p{1.6cm}p{2.3cm}p{1.7cm}p{1.2cm}}
\toprule
\textbf{Question Category} & \textbf{Definition} & \textbf{Examples} & \textbf{Visual Context} \\
\midrule
Visual fact extraction (text and graphics) &
Seeks information from text, symbols, labels, diagrams, or designs. &
"Tell me the text of the signs." &
\raisebox{-0.7\height}{\includegraphics[width=1.2cm]{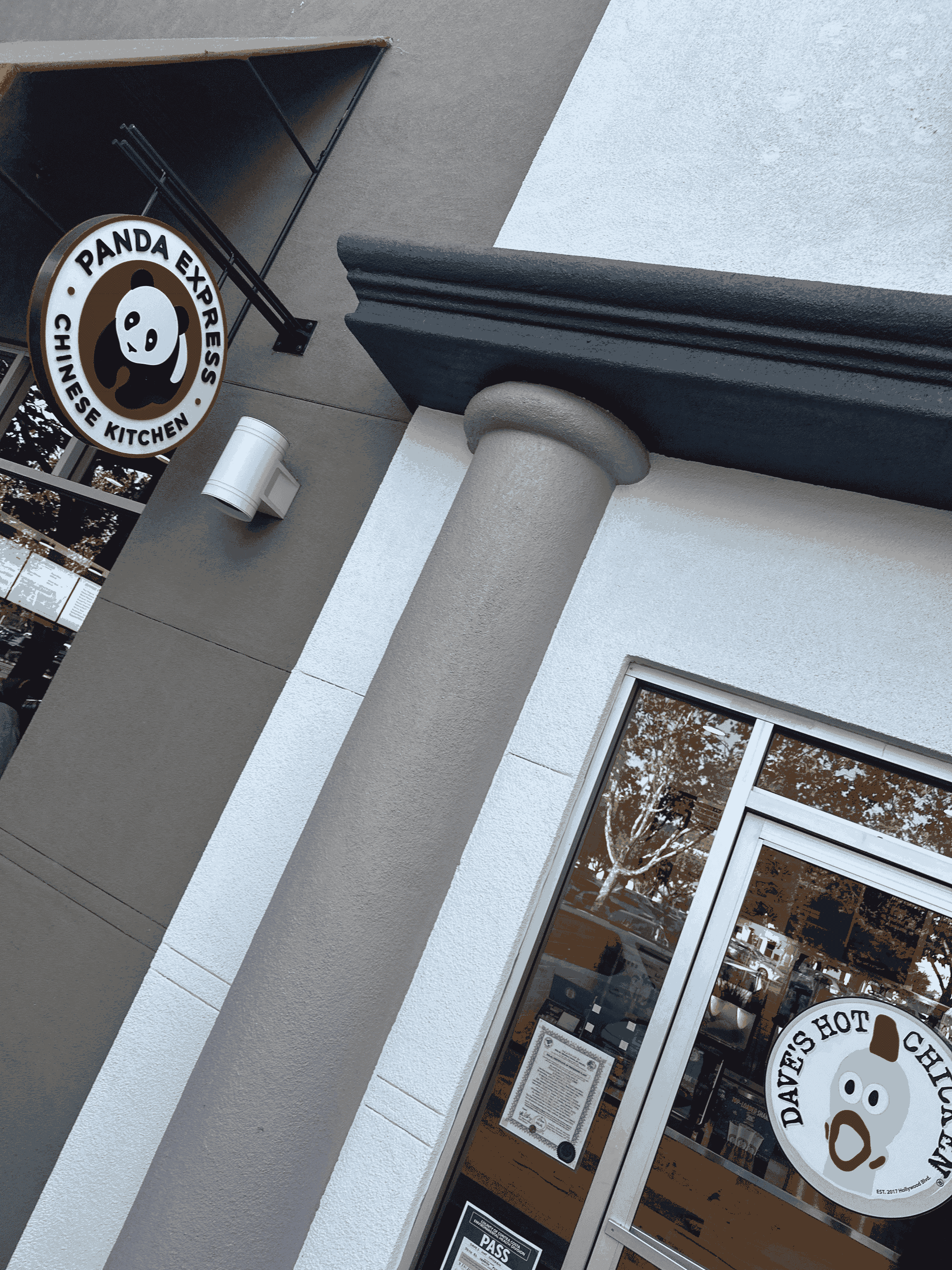}} \\

Visual fact extraction (quality) &
Asking about visual attributes from an entity (e.g., color, shape, appearance, size, or state.) &
"Does it have realistic features?" &
\raisebox{-0.7\height}{\includegraphics[width=1.2cm]{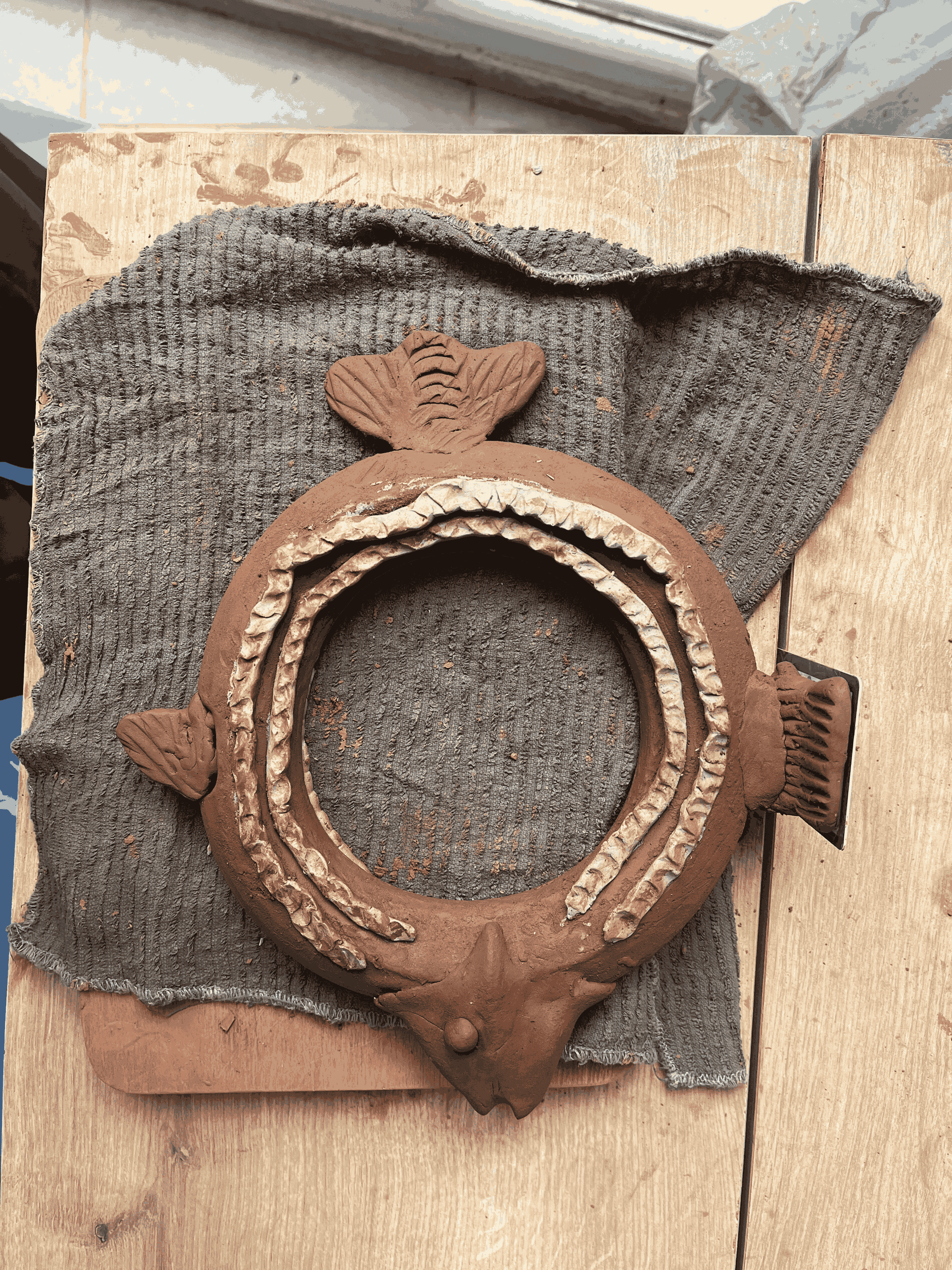}} \\

Visual fact extraction (unspecified) &
Asks for a general description without specifying what information is desired. &
"Describe my image on the screen." &
\raisebox{-0.7\height}{\includegraphics[width=1.2cm]{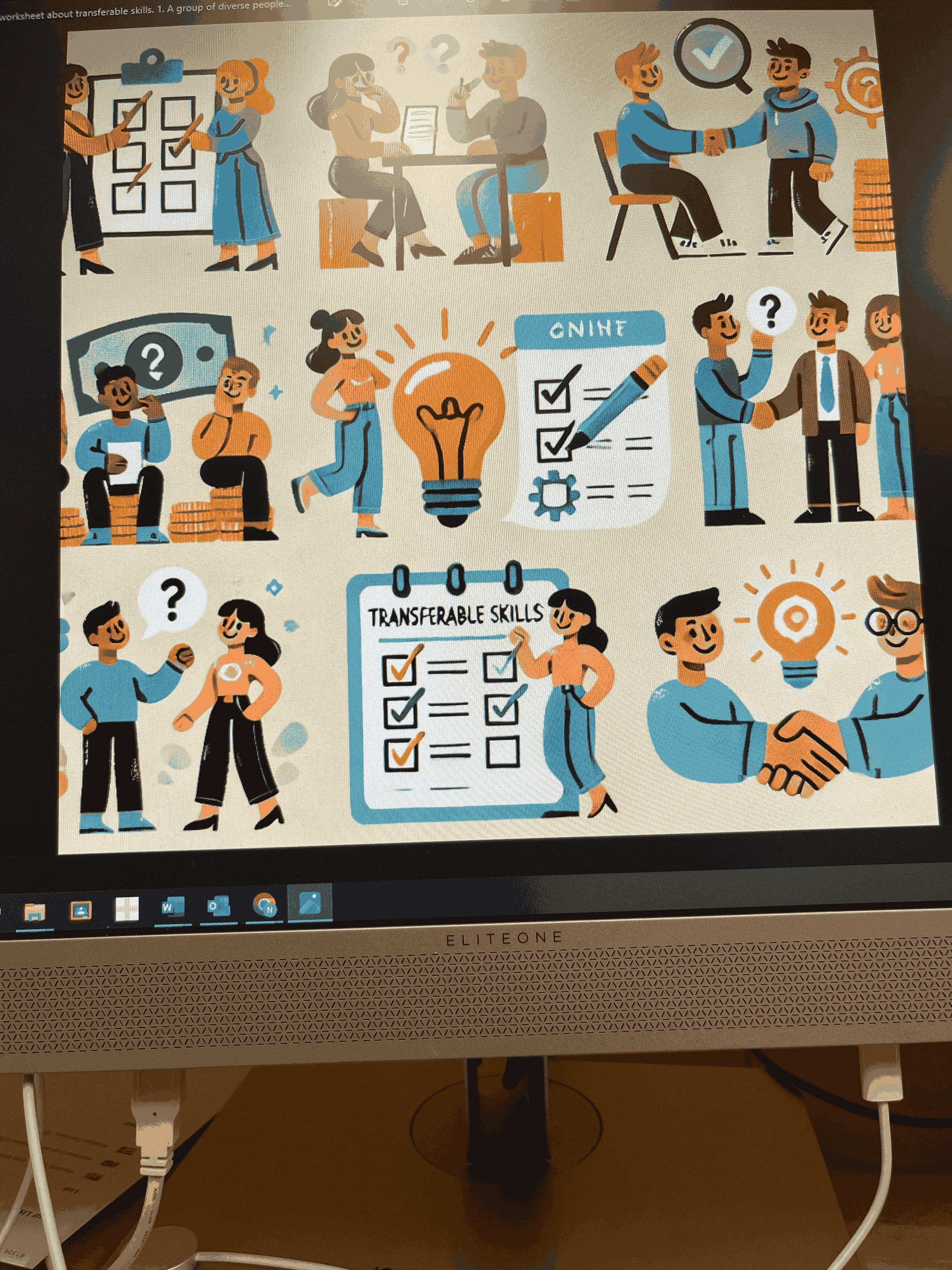}} \\

Verification \cite{chen2025fully} &
Confirm visual info based on a hypothesis the user proposes (e.g., object presence or identity). &
"Is there orange juice?" &
\raisebox{-0.7\height}{\includegraphics[width=1.2cm]{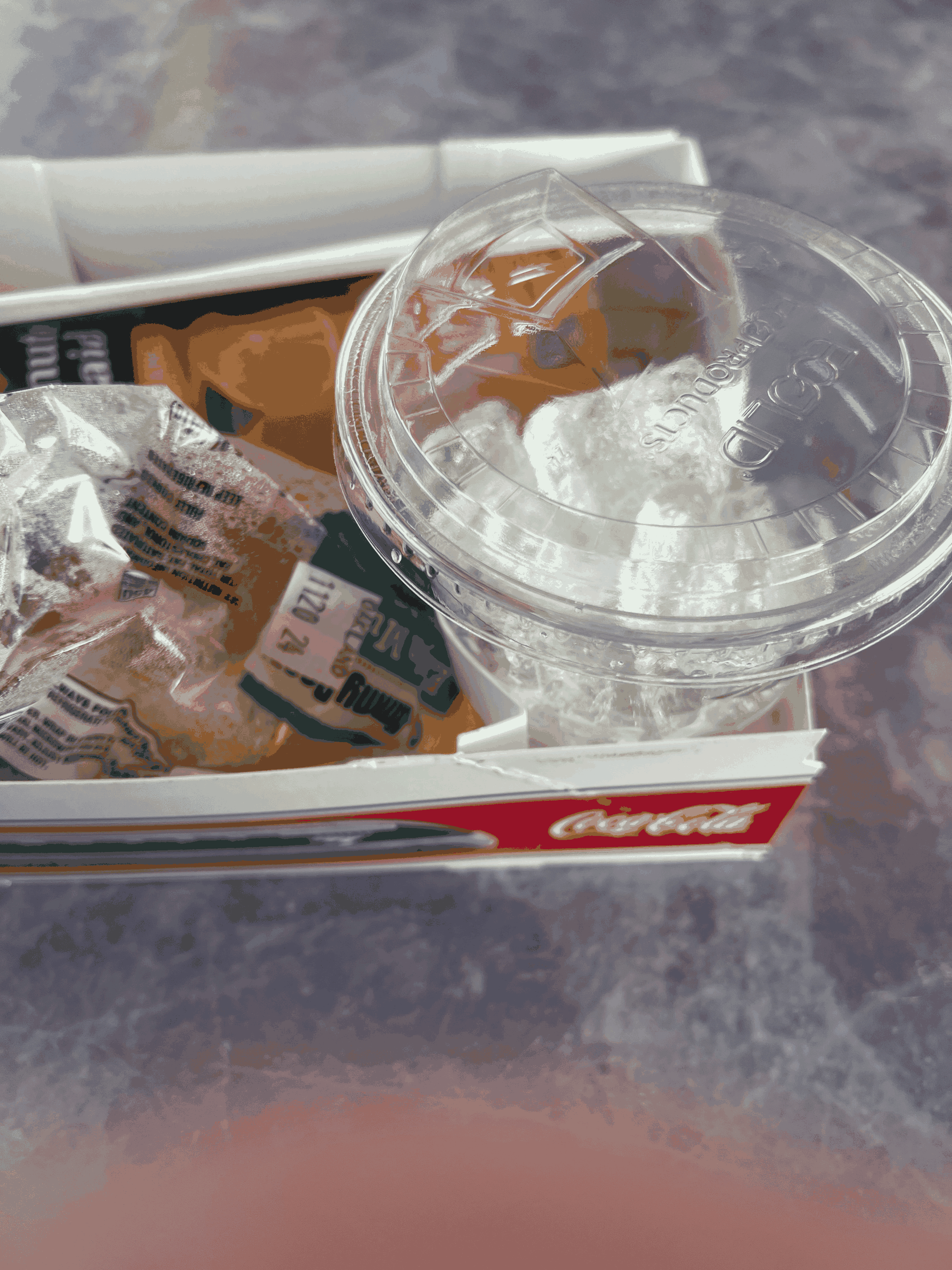}} \\

Identification \cite{chen2025fully} &
The expected answer to these questions is a named or group of named entities. &
"What kind of keyboard is it?" &
\raisebox{-0.7\height}{\includegraphics[width=1.2cm]{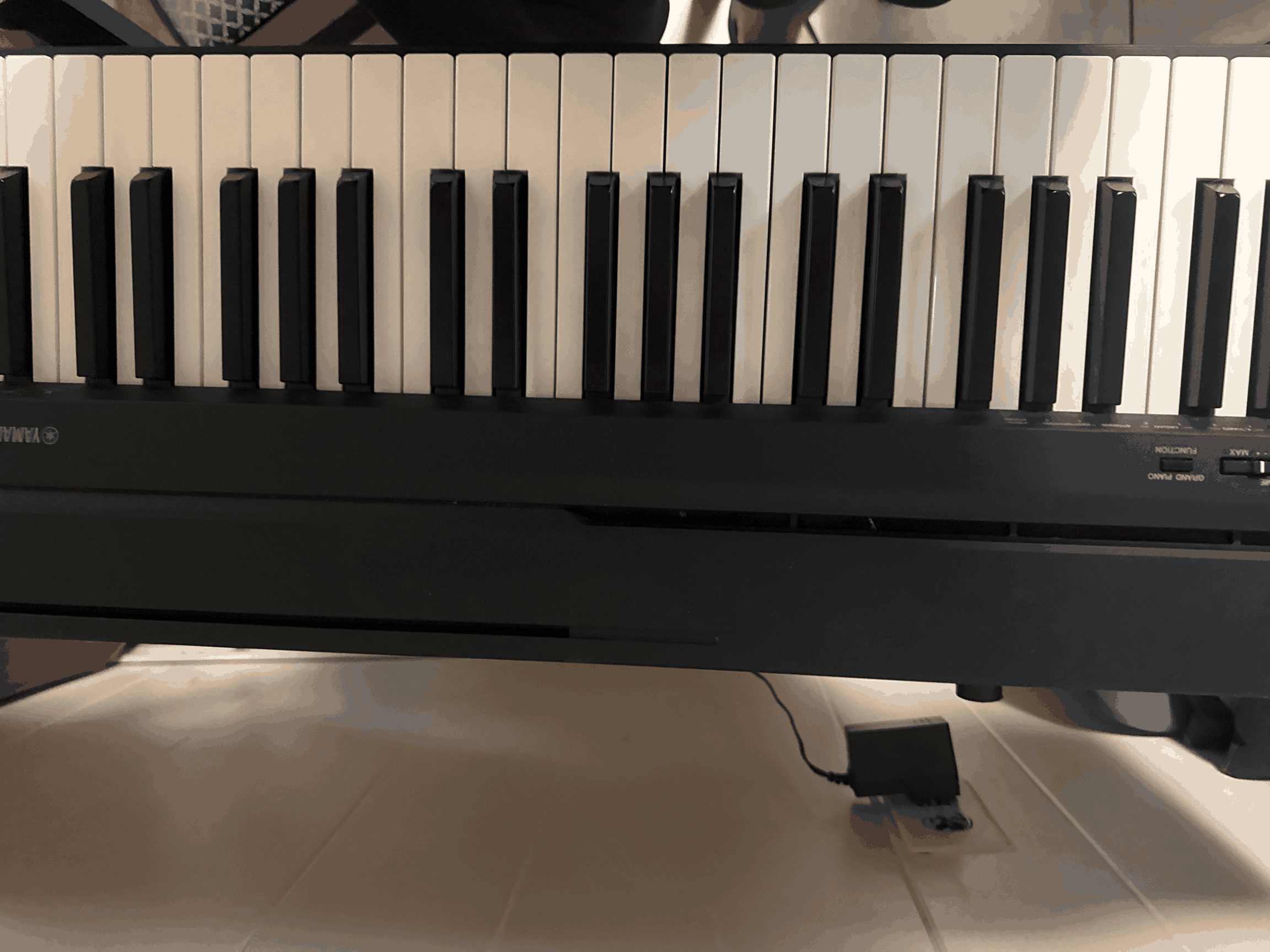}} \\

Instruction \cite{chen2025fully} &
Questions typically involve "how to do" inquiries. Grounded in the questioner's desire to learn steps or strategies that are known. &
"Is {[}blue face mask{]} on the right way?" &
\raisebox{-0.7\height}{\includegraphics[width=1.2cm]{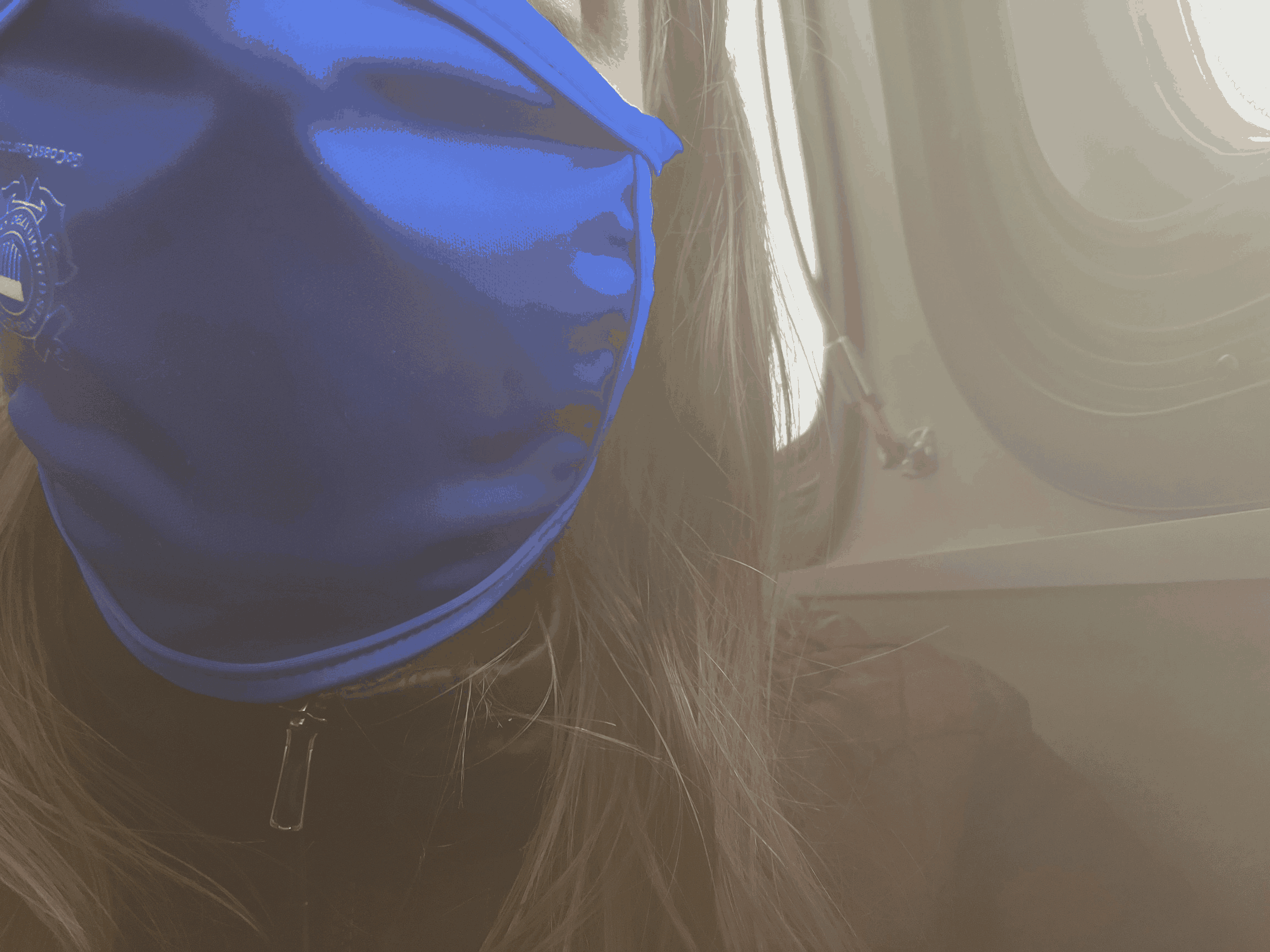}} \\

Localization &
Ask about object position relative to others or self. &
"In which position is root beer?" &
\raisebox{-0.7\height}{\includegraphics[width=1.2cm]{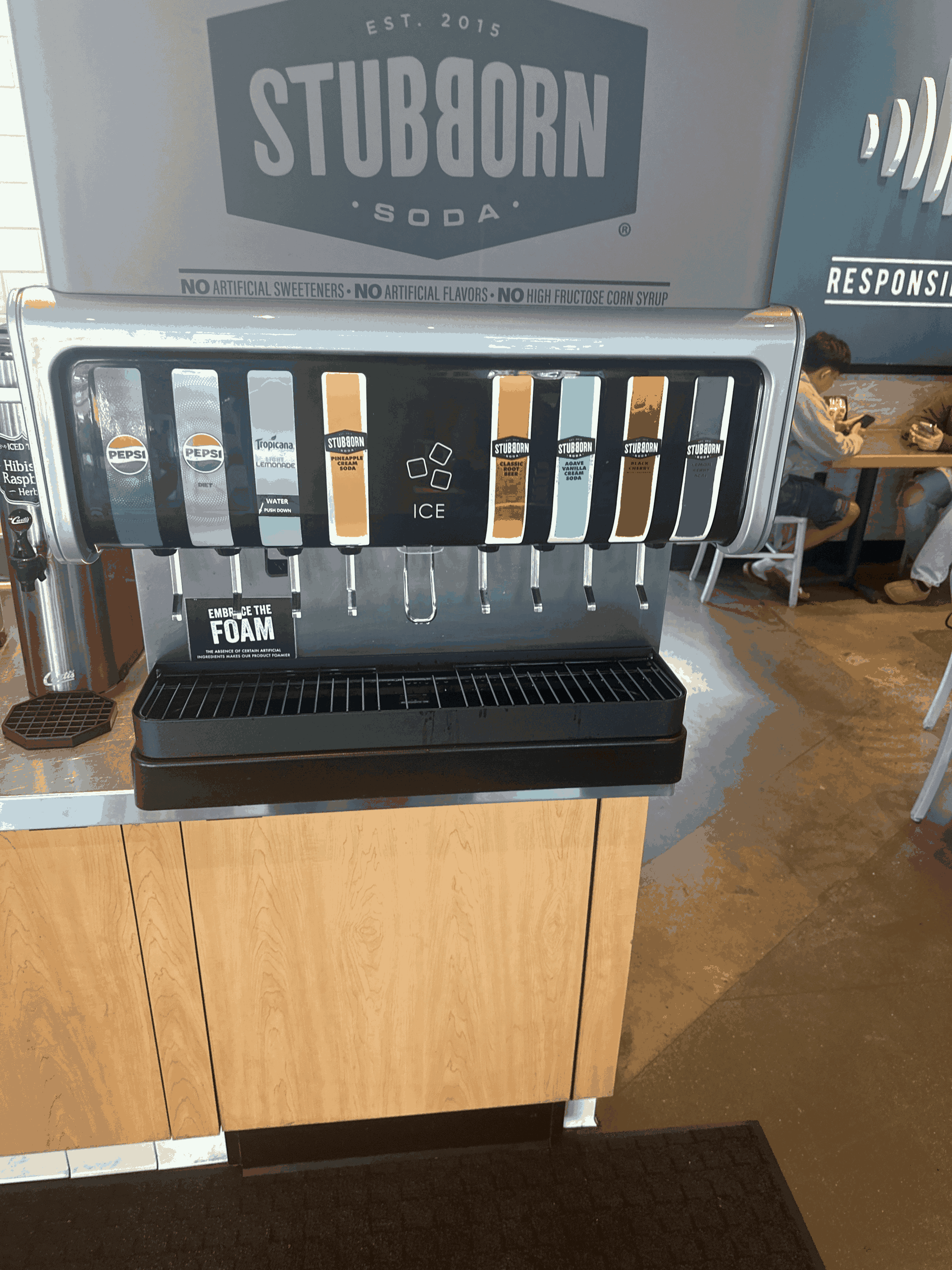}} \\
Advice \cite{chen2025fully} &
Questions where the user seeks personalized guidance on a specific topic. Users expect subjective recommendations. &
"Will this supplement help me?" &
\raisebox{-0.7\height}{\includegraphics[width=1.2cm]{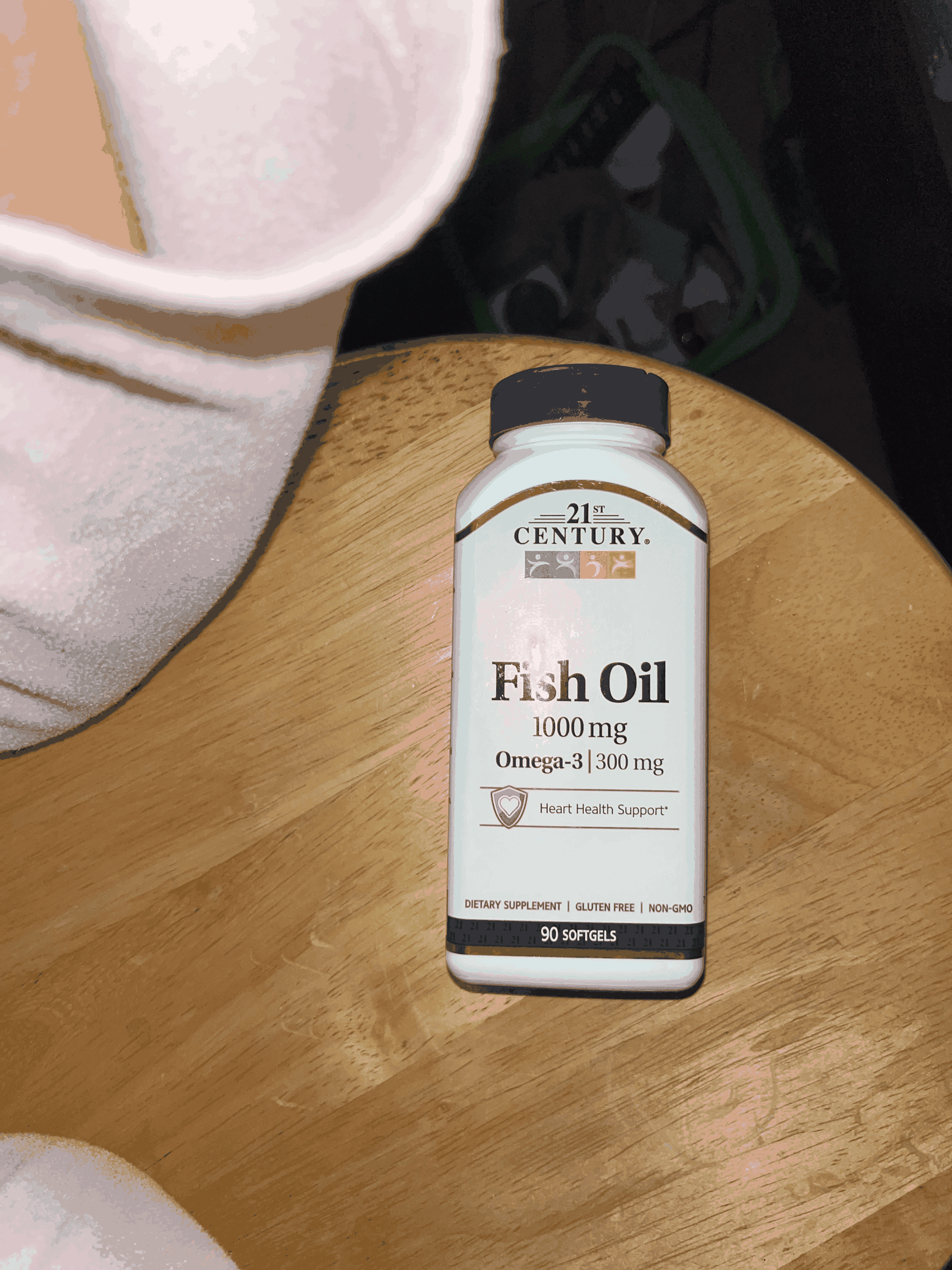}} \\
\bottomrule
\end{tabular}
\end{table}

\section{Location Categories}

\label{appendix:Location Categories}
In this section, we share examples of location categories where participants used our application. We determined these categories based on our codebook from our prior work \cite{gonzalez2024usecases}, and the information shared by the participant in the diary entry.
\begin{table}[htbp]
\caption{Categories of locations.}
\label{tab:location1}
\begin{tabular}{p{2.2cm}p{3.7cm}p{1.4cm}}
\toprule
\textbf{Location} & \textbf{Location Examples} & \textbf{Visual Context} \\
\midrule
Living Spaces &
Private homes, dormitories, and shared housing &
\raisebox{-0.7\height}{\includegraphics[width=1.2cm]{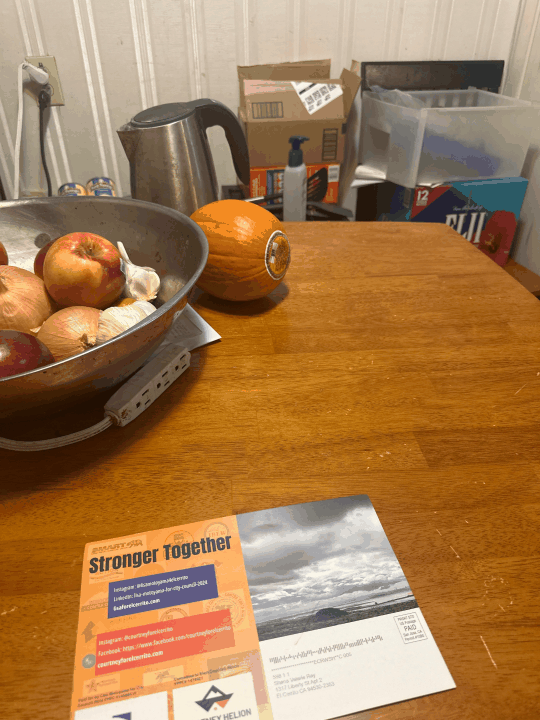}} \\
Recreational and Leisure Areas &
Cafes, restaurants, malls, museums, parks, and social spaces &
\raisebox{-0.7\height}{\includegraphics[width=1.2cm]{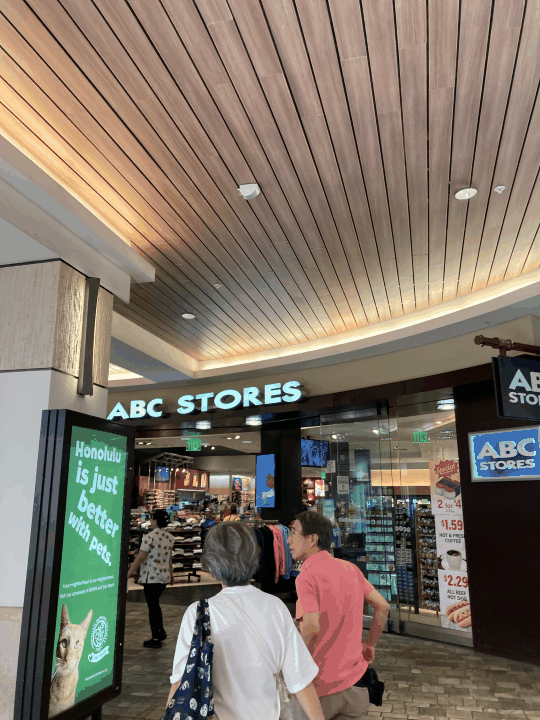}} \\
Work &
Offices, coworking spaces, construction sites, and corporate buildings &
\raisebox{-0.7\height}{\includegraphics[width=1.2cm]{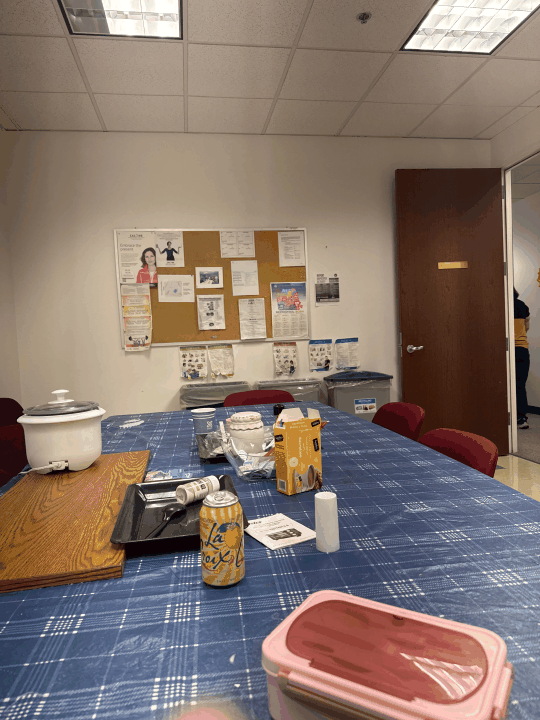}} \\
Transit Setting &
Subway stations, bus terminals, airports, and ferry docks &
\raisebox{-0.7\height}{\includegraphics[width=1.2cm]{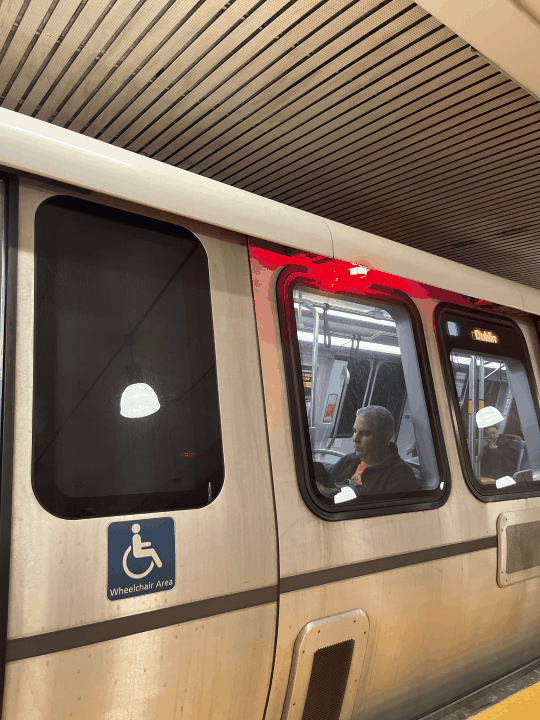}} \\
Unknown Location &
Unspecified or unclear environments based on visual context &
\raisebox{-0.7\height}{\includegraphics[width=1.2cm]{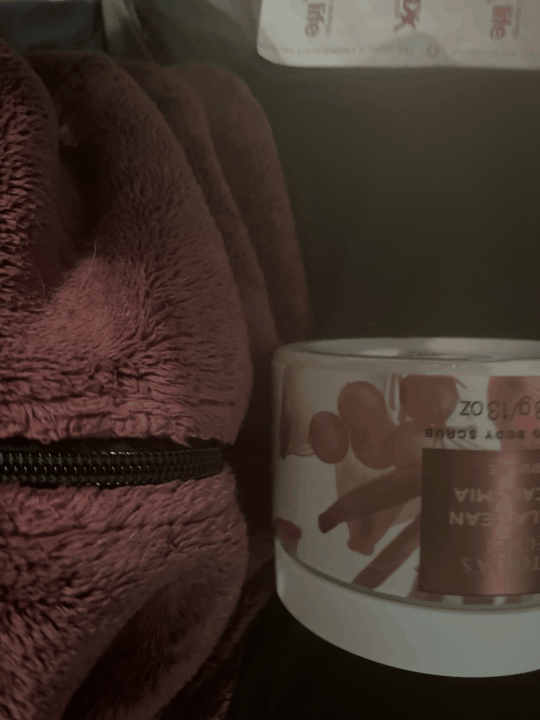}} \\
Retail Businesses &
Supermarkets, convenience stores, and specialty shops &
\raisebox{-0.7\height}{\includegraphics[width=1.2cm]{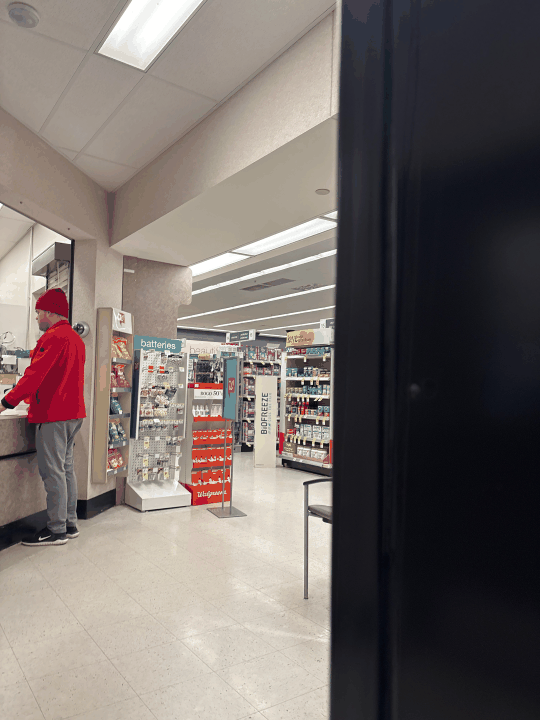}} \\
Educational and Spiritual Sites &
Classrooms, churches, temples, and places of worship &
\raisebox{-0.7\height}{\includegraphics[width=1.2cm]{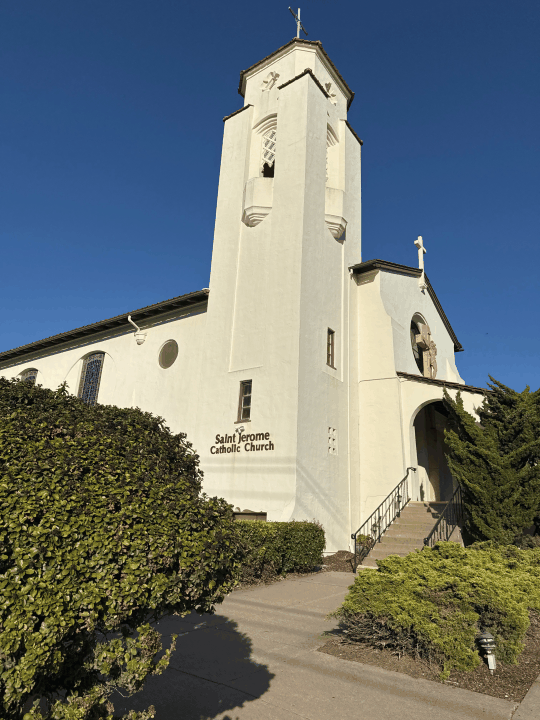}} \\
Healthcare Place &
Clinics, hospitals, therapy centers, and pharmacies &
\raisebox{-0.7\height}{\includegraphics[width=1.2cm]{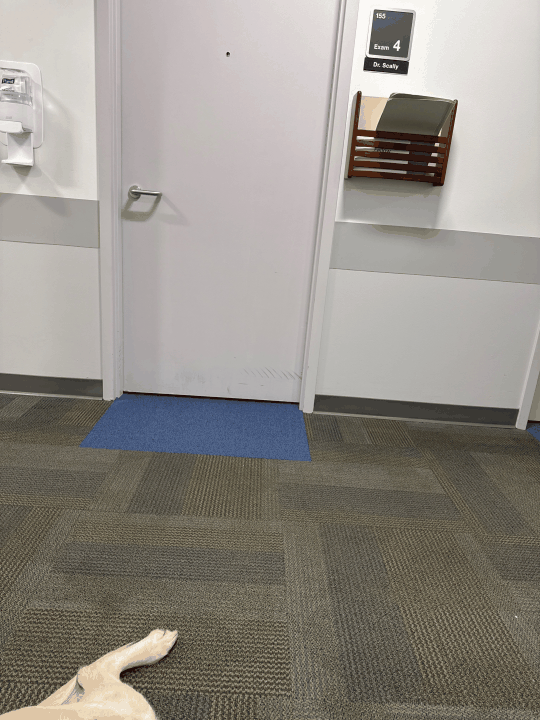}} \\
\bottomrule
\end{tabular}
\end{table}

\section{User Goal Categories}
\label{appendix:User Goal Categories}
In this section, we share examples of user goal categories. We determined these categories based on our codebook from our prior work \cite{gonzalez2024usecases}, and the information shared by the participant in the diary entry.
\begin{table}[htbp]
\caption{Categories of user goals (Part 1).}
\label{tab:UserGoals1}
\begin{tabular}{p{1.4cm}p{2.3cm}p{2.5cm}p{1.3cm}}
\toprule
\textbf{User Goal} & \textbf{Definition} & \textbf{Information Wanted Example} & \textbf{Photo Example} \\
\midrule
Identify Entity &
Trying to recognize an object, subcategory, or brand in the photo &
"To see if it would be able to guess or describe a breed that it could be, but it did not do that" &
\raisebox{-0.7\height}{\includegraphics[width=1.2cm]{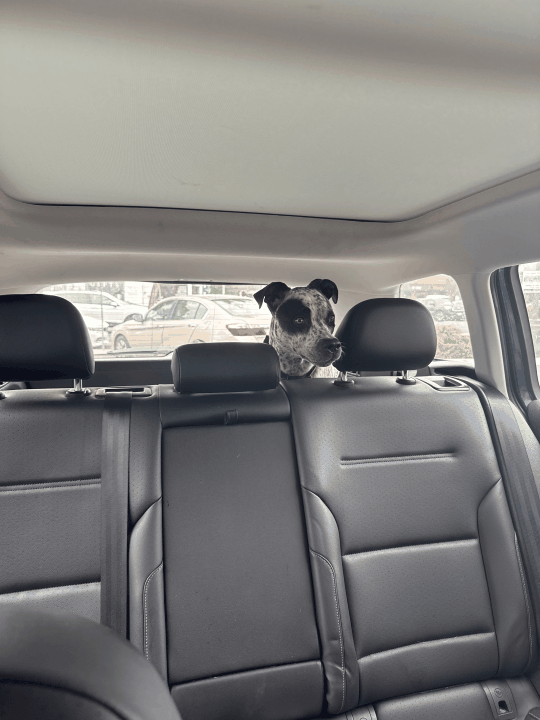}} \\

Read Text, Numbers, Graphics &
Wanting to read visible text, numbers, symbols, or graphical elements. &
"Type of salad and expiration date." &
\raisebox{-0.7\height}{\includegraphics[width=1.2cm]{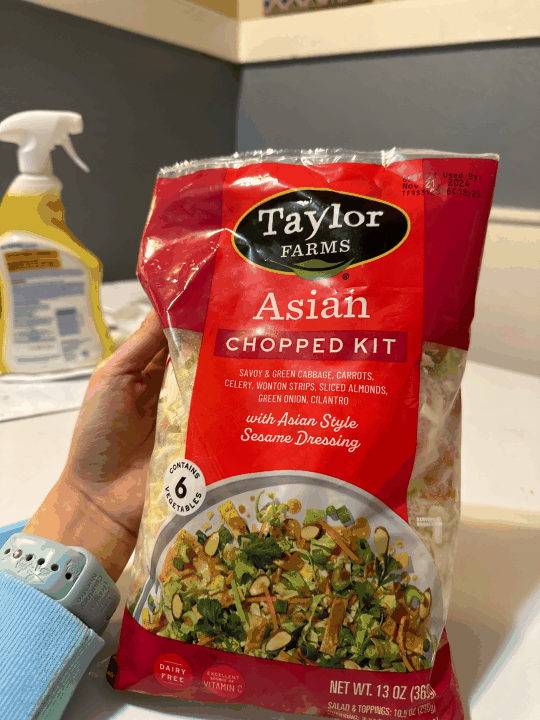}} \\

Receive Guidance &
Seeking advice, instruction, or assistance on what to do. &
"The preparation instructions." &
\raisebox{-0.7\height}{\includegraphics[width=1.2cm]{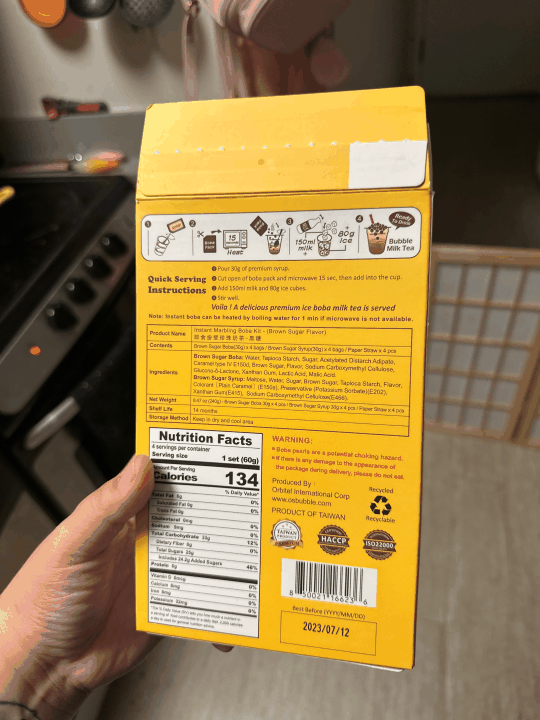}} \\

Determine Presence and Location &
Checking whether a specific person or object is present and identifying where it is within the frame. &
"I wanted the AI to explain what was on the ground. This was because I thought I might've dropped an AirPod." &
\raisebox{-0.7\height}{\includegraphics[width=1.2cm]{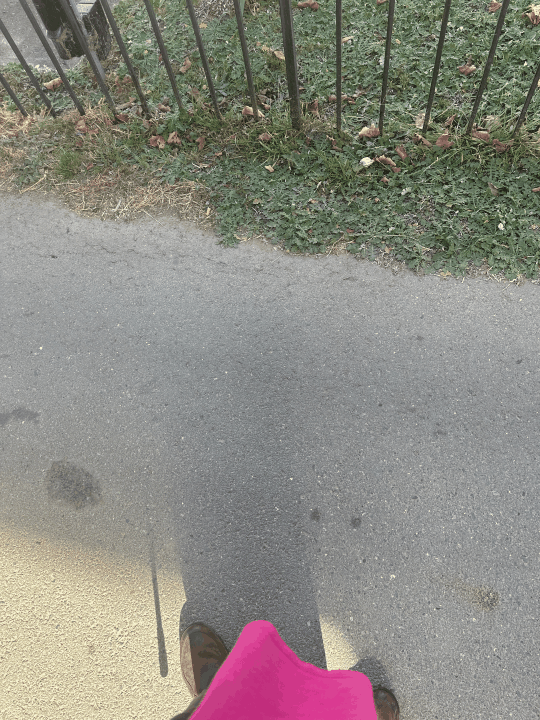}} \\
Personal Appearance &
Curious about their own appearance in the photo, such as hairstyle and clothing. &
"The color of the shirt." &
\raisebox{-0.7\height}{\includegraphics[width=1.2cm]{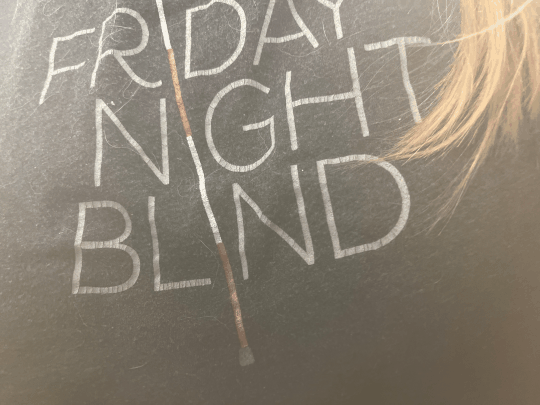}} \\

Describe Person &
Looking for details about a person in the photo, such as clothing, actions, expressions, and emotions. &
"A basic description of my mother." &
\raisebox{-0.7\height}{\includegraphics[width=1.2cm]{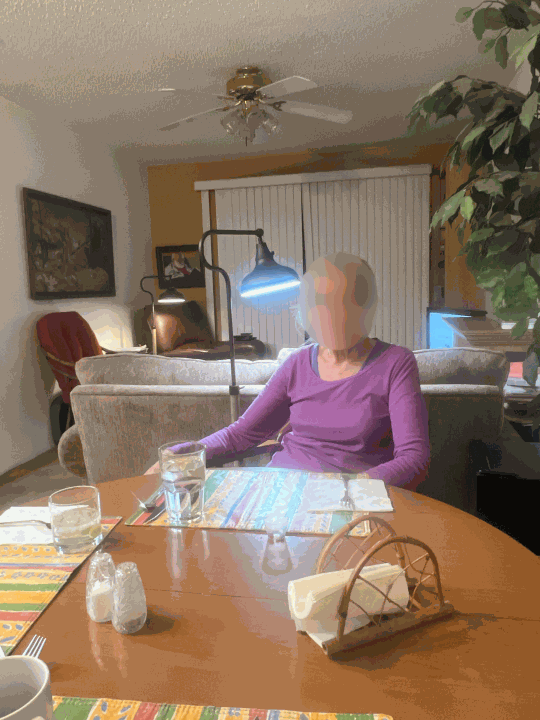}} \\

Learn Information about Subject &
Seeking contextual or background information about something depicted in the photo. &
"I wanted to know if it had Bluetooth connections." &
\raisebox{-0.7\height}{\includegraphics[width=1.2cm]{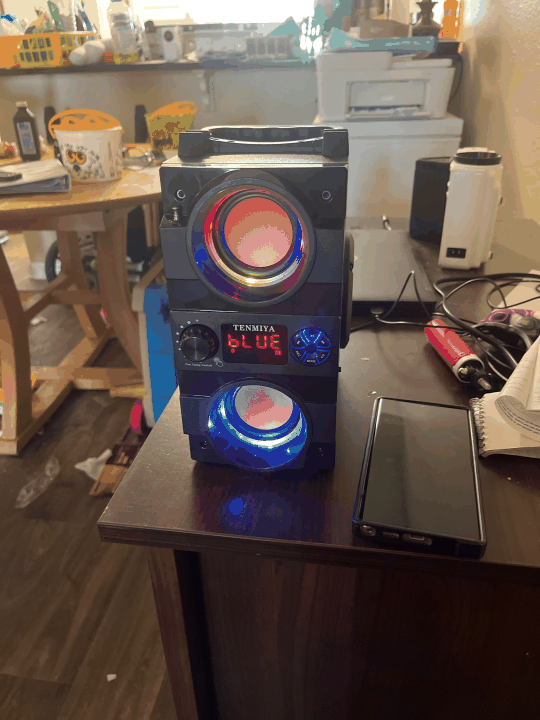}} \\
\bottomrule
\end{tabular}
\end{table}
\clearpage

\begin{table}[htbp]
\caption{Categories of user goals (Part 2).}
\label{tab:UserGoals2}
\begin{tabular}{p{1.2cm}p{2.3cm}p{2.5cm}p{1.3cm}}
\toprule
\textbf{User Goal} & \textbf{Definition} & \textbf{Information Wanted Example} & \textbf{Photo Example} \\
\midrule
Description of Subject &
Looking for a general description of the main item or focus in the photo. &
"Details of the picture on the postcard." &
\raisebox{-0.7\height}{\includegraphics[width=1.2cm]{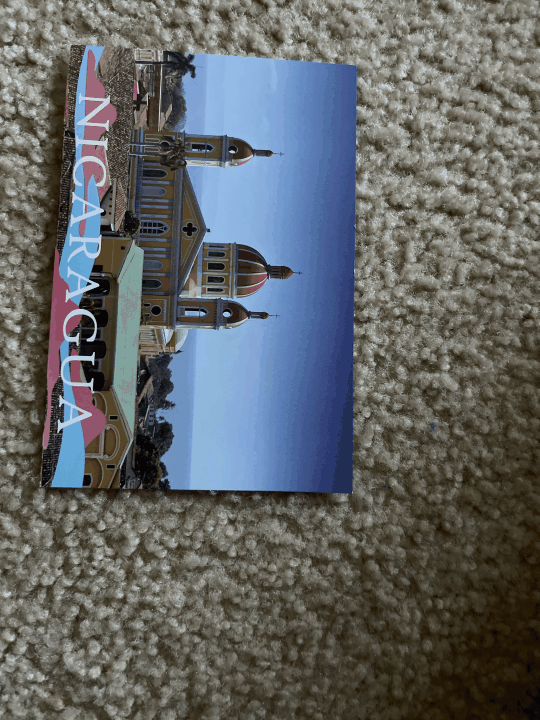}} \\

Learn Application &
Using the app to understand its functionality, test accuracy, or explore how it interprets content &
"I wanted the AI to describe accurately what was on the table in front of me." &
\raisebox{-0.7\height}{\includegraphics[width=1.2cm]{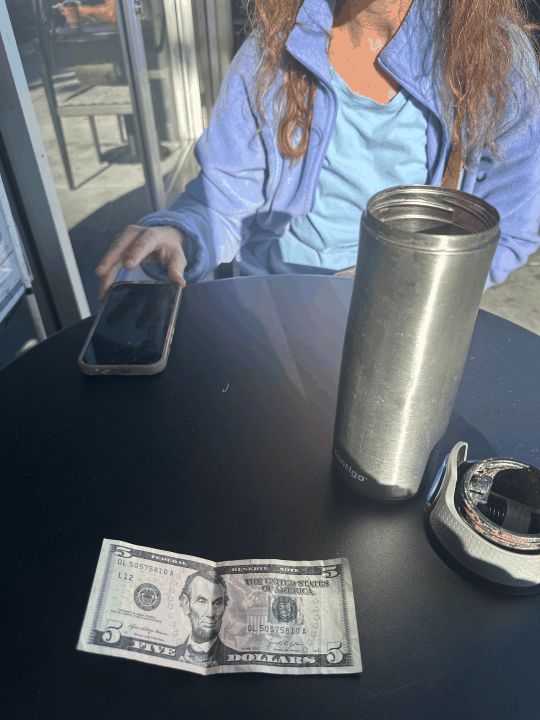}} \\

Build Understanding Scenery &
Looking for a description of the surrounding environment, context, or setting within the frame. &
"I just wanted to get an overall view of my surroundings." &
\raisebox{-0.7\height}{\includegraphics[width=1.2cm]{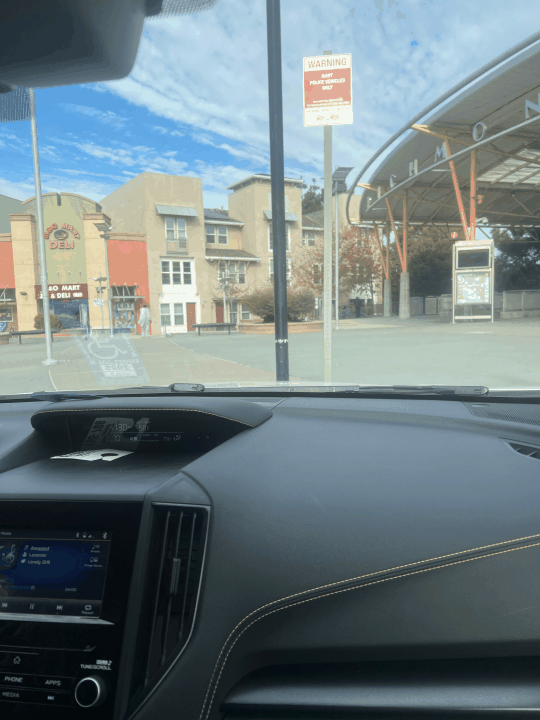}} \\

Describe Digital Content &
Looking for a description of items displayed on screens, such as apps, websites, or media content. &
"Want to know if the color bars look correct on the TV?" &
\raisebox{-0.7\height}{\includegraphics[width=1.2cm]{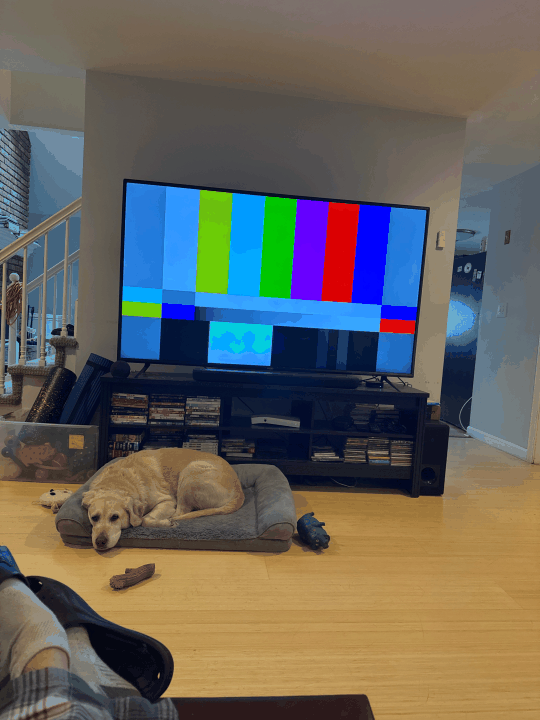}} \\

Navigation &
Asking for guidance or directions to move through or interact with a space. &
"Want to make sure I was at the right spot across the street." &
\raisebox{-0.7\height}{\includegraphics[width=1.2cm]{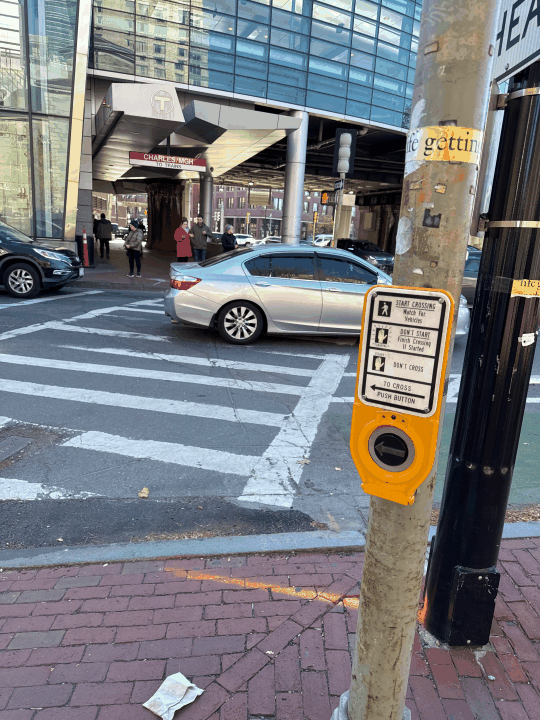}} \\

Identify Feature &
Looking for recognition of a specific feature or attribute (e.g. color, shape, label) of an item &
"Describe the color of the lipstick" &
\raisebox{-0.7\height}{\includegraphics[width=1.2cm]{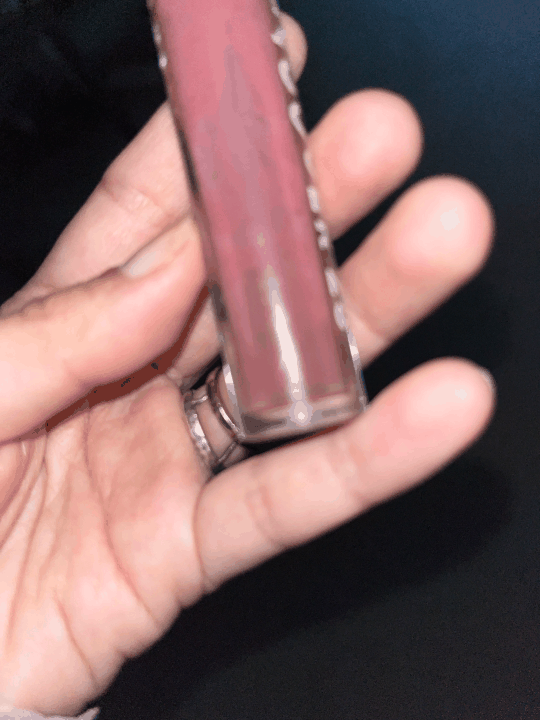}} \\

Understand Physical Interface &
Trying to identify physical buttons, ports, or other tactile features on devices. &
"To see how well it would describe where the parking indicator Was." &
\raisebox{-0.7\height}{\includegraphics[width=1.2cm]{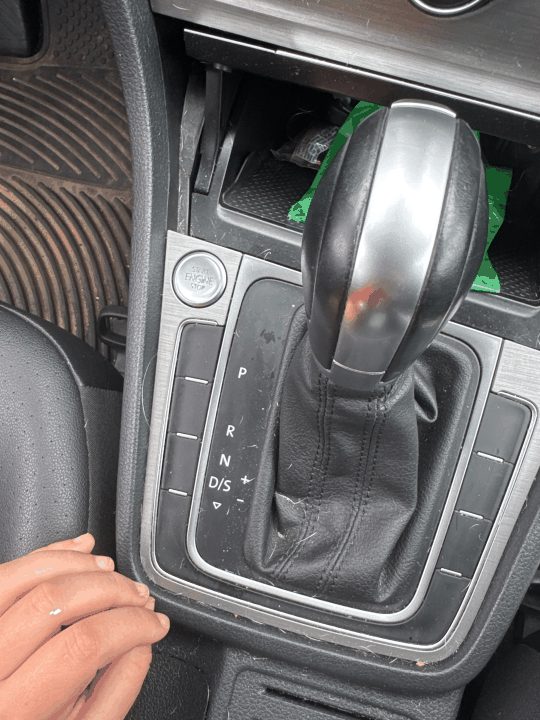}} \\

Unknown Goal &
Goal was unclear to the coders, or the user stated they had no specific goal in mind. &
N/A &
\raisebox{-0.7\height}{\includegraphics[width=1.2cm]{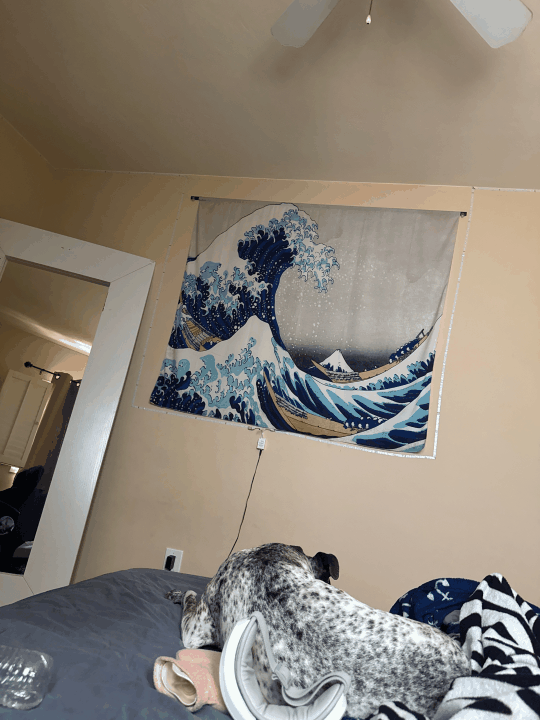}} \\
\bottomrule
\end{tabular}
\end{table}

\end{document}